\documentclass[preprintnumbers,showpacs,floats,twocolumn,prd,aps]{revtex4}
\usepackage{amssymb,amsmath}
\usepackage{epsfig,hyperref}
\usepackage{pgf,tikz}

\renewcommand{\vec}[1]{\mathbf{#1}}
\newcommand{\arXiv}[1]{\href{http://arxiv.org/abs/#1}{\texttt{arXiv:#1}}}

\begin{document}

\title{DEFROST: A New Code for Simulating Preheating after Inflation\bigskip\\
\url{http://www.sfu.ca/physics/cosmology/defrost/}}
\author{Andrei V. Frolov}\email{frolov@sfu.ca}
\affiliation{
  Department of Physics,
  Simon Fraser University\\
  8888 University Drive,
  Burnaby, BC Canada
  V5A 1S6
}
\date{September 28, 2008}

\begin{abstract}
  At the end of inflation, dynamical instability can rapidly deposit the
  energy of homogeneous cold inflaton into excitations of other fields. This
  process, known as preheating, is rather violent, inhomogeneous and
  non-linear, and has to be studied numerically. This paper presents a new
  code for simulating scalar field dynamics in expanding universe written for
  that purpose. Compared to available alternatives, it significantly improves
  both the speed and the accuracy of calculations, and is fully instrumented
  for 3D visualization. We reproduce previously published results on
  preheating in simple chaotic inflation models, and further investigate
  non-linear dynamics of the inflaton decay. Surprisingly, we find that the
  fields do not want to thermalize quite the way one would think. Instead of
  directly reaching equilibrium, the evolution appears to be stuck in a rather
  simple but quite inhomogeneous state. In particular, one-point distribution
  function of total energy density appears to be {\em universal} among various
  two-field preheating models, and is exceedingly well described by a
  lognormal distribution. It is tempting to attribute this state to scalar
  field turbulence.
\end{abstract}

\preprint{SCG-2008-02}
\maketitle

\section{Introduction}\label{sec:intro}

The idea of inflation is a cornerstone of the modern theory of the early
universe. According to inflationary paradigm, universe at early times
undergoes a period of rapid (quasi-exponential) expansion, which wipes the
initial state of the universe clean while seeding the primordial
inhomogeneities with quantum fluctuations generated during expansion
\cite{Linde:2005dd, Linde:2005ht}. While universe is inflating, all of its
energy sits in the homogeneous scalar field or condensate (known as inflaton),
which is in a vacuum-like state with little entropy or particle excitations.
But eventually inflation ends, and this energy has to be deposited into
excitations of other matter fields, starting the thermal history of the
universe with a hot big bang.

Decay of the inflaton can be very efficient if the fields experience dynamical
instability at the end of inflation; such a stage became known as preheating.
In most chaotic inflation models, oscillations of the inflaton field can cause
instability via parametric resonance \cite{Dolgov:1989us, Traschen:1990sw,
Kofman:1994rk, Shtanov:1994ce, Kofman:1995fi}. Although linear development of
this instability can be understood analytically \cite{Kofman:1997yn,
Greene:1997fu}, it might be chaotic \cite{Podolsky:2002qv}, and one needs to
resort to numerical simulations to investigate non-linear dynamics that soon
takes over \cite{Khlebnikov:1996mc,Podolsky:2005bw, Dufaux:2006ee,
Felder:2006cc}. In hybrid inflation models \cite{Linde:1993cn}, in addition to
parametric resonance \cite{GarciaBellido:1997wm}, one also has a tachyonic
instability associated with the symmetry breaking \cite{Felder:1998vq},
dynamics of which has been explored in \cite{Felder:2000hj, Felder:2001kt,
GarciaBellido:2002aj}.

Non-equilibrium dynamics of preheating can lead to a multitude of interesting
phenomena. Some of the topics discussed in the literature are formation of
topological defects \cite{Tkachev:1998dc, Battye:1998xe}, production of
various particles (with applications to baryo- and leptogenesis)
\cite{GarciaBellido:1999sv, GarciaBellido:2000dc, GarciaBellido:2001cb,
GarciaBellido:2003wd}, possibility of primordial black hole formation
\cite{Suyama:2004mz, Suyama:2006sr}, generation of primordial magnetic fields
\cite{DiazGil:2007dy, DiazGil:2008tf}, and production of stochastic
gravitational wave background \cite{Easther:2006gt, Easther:2006vd,
GarciaBellido:2007dg, GarciaBellido:2007af, Dufaux:2007pt, Caprini:2007xq}.
Due to difficulties of dealing with non-linear evolution equations, most of
these studies rely on numerical simulations.

This paper describes a new code for simulating non-linear scalar field
dynamics in expanding universe developed to study preheating, called
DEFROST, and the first results obtained with it. There are other codes
available for this purpose, most notably LATTICEEASY by Gary Felder and Igor
Tkachev \cite{Felder:2000hq}, and its parallel version CLUSTEREASY
\cite{Felder:2007nz}. Through the use of more advanced algorithms and careful
optimization, the new code significantly improves both accuracy and
performance achievable in simulations of preheating. An important design goal
has been the ease of visualization and analysis of the results, which is all
too important for understanding dynamics of complex systems.

We present results on preheating in a simple two-field chaotic inflation model
with massive inflaton decaying into another scalar field via quartic coupling
\cite{Felder:2006cc}. Through our simulations, a new and somewhat simpler
picture of the late stages of preheating emerges. After initial transient when
instability develops, bubbles form and then break-up, the matter distribution
soon arranges itself in a clumpy state which persists with little changes for
a long time. One-point probability distribution function of total energy
density in this state appears to be {\em universally lognormal} among various
two-field preheating models. It is tempting to attribute this to relativistic
turbulence \cite{Nordlund:1998wj, Micha:2002ey, Micha:2004bv}. We also see
some evidence that the structure formed during preheating continues to grow in
size on a much longer timescale. Somehow, this picture reminds one of large scale
structure formation \cite{Bardeen:1985tr, Bernardeau:1994aq, Bond:1995yt}.

This paper is organized in the following way: In Section~\ref{sec:eom}, we
introduce scalar field models of preheating, derive equations of motion, and
discuss physical approximations we use. Section~\ref{sec:solver} describes the
detailed implementation of numerical evolution algorithm. Initial conditions
including quantum fluctuations of the fields produced during inflation are
discussed in Section~\ref{sec:ic}, with particular attention paid to
implementation of Gaussian random field generator. We briefly recount the
theory of preheating via broad parametric resonance at the end of chaotic
inflation in Section~\ref{sec:chaotic}, and present our simulations in
Section~\ref{sec:results}. We conclude by summarizing our main results in
Section~\ref{sec:concl}.

\section{Equations of Motion}\label{sec:eom}

As our baseline model of reheating we will take a system of $N$ scalar fields
$\{\phi^i\}$, minimally coupled to gravity and interacting through some
(non-linear) potential $V(\phi^i)$ as described by the action
\begin{equation}\label{eq:action}
  S = \frac{1}{16\pi G} \int\left\{R - g^{\mu\nu}\delta_{ij} \phi^i_{;\mu}\phi^j_{;\nu} - 2V(\phi^i)\right\}\, \sqrt{-g}\, d^4x.
\end{equation}
The above action can be modified to describe more complicated geometrical
quantities (like a vector field, for example) instead of a real scalar
multiplet $\{\phi^i\}$. Ultimately what we care about is only field equations
of motion and their gravitational effects, not actual gauge symmetries.
Variation of the action (\ref{eq:action}) with respect to the metric gives
Einstein equation with a stress-energy tensor
\begin{equation}\label{eq:T}
  T_{\mu\nu} = \delta_{ij} \phi^i_{;\mu}\phi^j_{;\nu} -
    \left[\frac{1}{2} (g^{\alpha\beta}\delta_{ij} \phi^i_{;\alpha}\phi^j_{;\beta}) + V(\phi^i)\right] g_{\mu\nu},
\end{equation}
while variation with respect to the field $\phi^i$ gives equation of motion for the field
\begin{equation}\label{eq:box}
  \Box \phi^i \equiv g^{\mu\nu}\phi^i_{;\mu\nu} = \frac{\partial V}{\partial\phi^i}.
\end{equation}
Although in principle it is possible to solve the complete system of Einstein
and scalar field equations numerically (see \cite{York:1971hw, Shibata:1995we,
Baumgarte:1998te, Pretorius:2004jg} for formulation and some approaches), in
practice, it is not such an easy thing to do. Einstein equation solvers in
$3+1$ dimensions are complex to implement, very expensive to run, and, despite
marked improvement in the recent years \cite{Pretorius:2004jg}, still might
have issues with numerical stability.

Fortunately for us, we do not have to solve the full Einstein equations.
Although preheating is a rather violent process, and stress-energy tensor
becomes very inhomogeneous due to non-linear field dynamics, the smallness of
gravitational coupling constant assures that gravitational backreaction of these
inhomogeneities is rather small (at $10^{-3}$ level in the simulations
presented in this paper). Thus we are going to treat the scalar field evolution as if
it was happening in a homogeneous flat Friedman-Robertson-Walker spacetime
\begin{equation}\label{eq:frw}
  ds^2 = -dt^2 + a^2(t)\, d\vec{x}^2,
\end{equation}
and calculate the inhomogeneous gravitational field due to matter distribution
with stress-energy tensor (\ref{eq:T}) using linear perturbation theory \cite{Bardeen:1980kt}.
We will ignore back-reaction of metric perturbations on the scalar field evolution.

With these simplifying assumptions, the problem becomes much more tractable:
we just need to solve a system of coupled scalar field equations of motion
(\ref{eq:box}), which in spacetime with metric (\ref{eq:frw}) become
\begin{equation}\label{eq:phi}
  \ddot{\phi}^i + 3H\dot{\phi}^i - \frac{\Delta}{a^2}\, \phi^i + \frac{\partial V}{\partial\phi^i} = 0.
\end{equation}
Here and later, dot denotes the derivative with respect to time $t$, and
spatial differential operators (gradient $\nabla$ and Laplacian $\Delta$) are
taken with respect to comoving coordinates and three-dimensional flat metric.
The expansion rate $H \equiv \dot{a}/a$ and acceleration $\ddot{a}$ are
determined by averaged Einstein equations
\begin{equation}\label{eq:H}
  H^2 = \frac{\langle\rho\rangle}{3}, \hspace{1em}
  \frac{\ddot{a}}{a} = - \frac{1}{6}\, \langle\rho + 3p\rangle,
\end{equation}
where energy density $\rho$ and isotropic pressure $p$ are components of the
stress energy tensor (\ref{eq:T}) given by
\begin{eqnarray}
  && \rho \equiv -T^t_t =
    \sum_i \frac{(\dot{\phi}^i)^2}{2} + \sum_i \frac{(\nabla\phi^i)^2}{2a^2} + V(\phi^i), \label{eq:rho}\\
  && p \equiv \frac{1}{3}\, T^a_a =
    \sum_i \frac{(\dot{\phi}^i)^2}{2} - \sum_i \frac{(\nabla\phi^i)^2}{6a^2} - V(\phi^i), \label{eq:p}
\end{eqnarray}
and averages are taken over the whole simulation volume.

To solve equation (\ref{eq:phi}) numerically, one needs to know the Hubble
parameter value $H$. Rather than attempting to resolve the constraint equation
(\ref{eq:H}) at every time step (which would result in an implicit evolution
scheme), it is faster and more convenient to use the evolution equation
\begin{equation}\label{eq:H'}
  \dot{H} = -H^2 - \frac{1}{6}\, \langle\rho + 3p\rangle = -\frac{1}{2}\, \langle\rho+p\rangle
\end{equation}
to evaluate its value in the future. This is what is done in LATTICEEASY.
However, we have one more trick up our sleeves: a disproportionately huge gain
in numerical accuracy can be realized by evolving the Hubble length $L \equiv
1/H$ instead of the Hubble parameter $H$ by using
\begin{equation}\label{eq:L}
  \dot{L} \equiv -\frac{\dot{H}}{H^2} = 1 + \frac{L^2}{6} \langle\rho + 3p\rangle.
\end{equation}
For constant equation of state $p=w\rho$, Hubble parameter evolves as $H
\propto 1/t$, while Hubble length evolves as $L \propto t$. The latter
variable has vanishing second (and higher) derivatives and correspondingly
smaller truncation error when discretized to second order. As an added bonus,
the spatial gradients (which are the single most expensive thing to calculate)
cancel out when taken in $\rho+3p$ combination, and do not enter evolution
equation in the form (\ref{eq:L}).

Once the field equations of motion (\ref{eq:phi}) are solved, we can evaluate
all the components of stress-energy tensor (\ref{eq:T}), in particular energy
density (\ref{eq:rho}) and pressure (\ref{eq:p}), as well as calculate linear
metric perturbations they create in a homogeneous spacetime (\ref{eq:frw}). In
this paper, we will focus on scalar perturbations, behaviour of which during
preheating have not been widely studied yet. In the longitudinal gauge,
perturbed metric can be written as
\begin{equation}\label{eq:pert}
  ds^2 = -(1+2\Phi) dt^2 + a^2 (1-2\Psi) d\vec{x}^2.
\end{equation}
The two scalar gravitational potentials $\Phi$ and $\Psi$ are equal in the
absence of anisotropic stress. This is in general not the case for the scalar
field models, and can be expected to hold only approximately and in the
average sense. Both gravitational potentials $\Phi$ and $\Psi$ are
non-dynamical, being solutions of the Poisson-like equations. In particular,
the equation for $\Psi$ is
\begin{equation}\label{eq:Psi}
  \frac{\Delta}{a^2}\,\Psi = \frac{\rho}{2}.
\end{equation}
Strictly speaking, what should stand in the right hand side is not the density
$\rho$, but the gauge-invariant density variable $\rho_m$, which has some
extra terms in it \cite{Bardeen:1980kt}. At this stage of the code
development, I will not make that distinction and simply ignore the missing
terms (which usually works well on sub-horizon scales), along with anisotropic
stress contribution to potential $\Phi$. Full support of gauge-invariant
perturbations including tensor and vector modes is planned for the future code
release.

\section{PDE Solver Implementation}\label{sec:solver}

Scalar field equations of motion (\ref{eq:phi}) are coupled non-linear partial
differential equations, and have to be solved numerically. Fortunately, all the
non-linearity comes from the potential term only; the differential operator
itself is simple, homogeneous and hyperbolic, which makes the numerical
solution quite straightforward. For the solver, we adopt a second-order
accurate finite difference scheme based on leapfrog algorithm.

The scalar field values $\phi^i$ are discretized on three-dimensional cubic $n
\times n \times n$ grid in comoving coordinates with uniform spacing $dx$ and
periodic boundary conditions. Since evolution equations (\ref{eq:phi}) are
second order, values of the fields on two consecutive time slices are required
to advance to the next one. We will denote the previous, current, and next
time slices by indices \texttt{dn}, \texttt{hr}, and \texttt{up}
correspondingly. The time derivatives of a quantity $X$ are discretized to
second order as
\begin{equation}\label{eq:disc}
  \dot{X} = \frac{X_{\text{up}} - X_{\text{dn}}}{2\, dt}, \hspace{1em}
  \ddot{X} = \frac{X_{\text{up}} - 2 X_{\text{hr}}+ X_{\text{dn}}}{(dt)^2}.
\end{equation}

\begin{figure}\large
\begin{tikzpicture}[x=6em,y=6em,z={(2em,1.5em)}]
  \tikzstyle{rank0}=[circle,draw=black,fill=blue!70,thick]
  \tikzstyle{rank1}=[circle,draw=black,fill=blue!50,thick]
  \tikzstyle{rank2}=[circle,draw=black,fill=blue!30,thick]
  \tikzstyle{rank3}=[circle,draw=black,fill=blue!10,thick]
  
  \node (000) at ( 0, 0, 0) [rank0] {$c_0$};
  \node (100) at ( 1, 0, 0) [rank1] {$c_1$};
  \node (010) at ( 0, 1, 0) [rank1] {$c_1$};
  \node (001) at ( 0, 0, 1) [rank1] {$c_1$};
  \node (I00) at (-1, 0, 0) [rank1] {$c_1$};
  \node (0I0) at ( 0,-1, 0) [rank1] {$c_1$};
  \node (00I) at ( 0, 0,-1) [rank1] {$c_1$};
  \node (110) at ( 1, 1, 0) [rank2] {$c_2$};
  \node (101) at ( 1, 0, 1) [rank2] {$c_2$};
  \node (011) at ( 0, 1, 1) [rank2] {$c_2$};
  \node (I10) at (-1, 1, 0) [rank2] {$c_2$};
  \node (I01) at (-1, 0, 1) [rank2] {$c_2$};
  \node (0I1) at ( 0,-1, 1) [rank2] {$c_2$};
  \node (1I0) at ( 1,-1, 0) [rank2] {$c_2$};
  \node (10I) at ( 1, 0,-1) [rank2] {$c_2$};
  \node (01I) at ( 0, 1,-1) [rank2] {$c_2$};
  \node (II0) at (-1,-1, 0) [rank2] {$c_2$};
  \node (I0I) at (-1, 0,-1) [rank2] {$c_2$};
  \node (0II) at ( 0,-1,-1) [rank2] {$c_2$};
  \node (111) at ( 1, 1, 1) [rank3] {$c_3$};
  \node (11I) at ( 1, 1,-1) [rank3] {$c_3$};
  \node (1I1) at ( 1,-1, 1) [rank3] {$c_3$};
  \node (1II) at ( 1,-1,-1) [rank3] {$c_3$};
  \node (I11) at (-1, 1, 1) [rank3] {$c_3$};
  \node (I1I) at (-1, 1,-1) [rank3] {$c_3$};
  \node (II1) at (-1,-1, 1) [rank3] {$c_3$};
  \node (III) at (-1,-1,-1) [rank3] {$c_3$};
  
  \draw (III) to (II0) to (II1);
  \draw (I0I) to (I00) to (I01);
  \draw (I1I) to (I10) to (I11);
  \draw (0II) to (0I0) to (0I1);
  \draw (00I) to (000) to (001);
  \draw (01I) to (010) to (011);
  \draw (1II) to (1I0) to (1I1);
  \draw (10I) to (100) to (101);
  \draw (11I) to (110) to (111);
  \draw (III) to (I0I) to (I1I);
  \draw (II0) to (I00) to (I10);
  \draw (II1) to (I01) to (I11);
  \draw (0II) to (00I) to (01I);
  \draw (0I0) to (000) to (010);
  \draw (0I1) to (001) to (011);
  \draw (1II) to (10I) to (11I);
  \draw (1I0) to (100) to (110);
  \draw (1I1) to (101) to (111);
  \draw (III) to (0II) to (1II);
  \draw (I0I) to (00I) to (10I);
  \draw (I1I) to (01I) to (11I);
  \draw (II0) to (0I0) to (1I0);
  \draw (I00) to (000) to (100);
  \draw (I10) to (010) to (110);
  \draw (II1) to (0I1) to (1I1);
  \draw (I01) to (001) to (101);
  \draw (I11) to (011) to (111);
\end{tikzpicture}
\caption{Three-dimensional spatial discretization stencil.}
\label{fig:disc}
\end{figure}
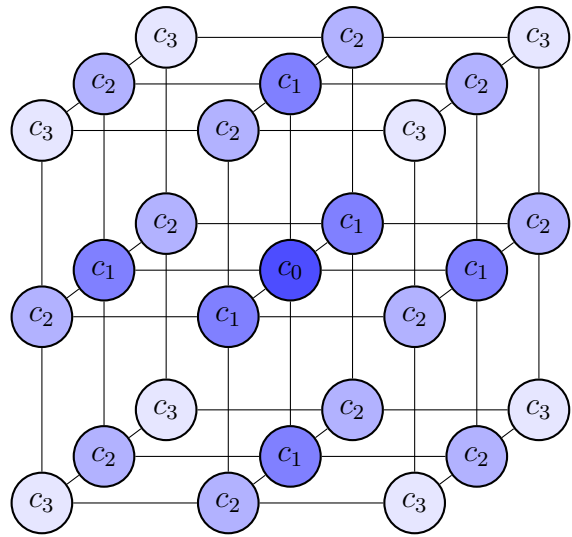

\begin{table}
  \begin{tabular}{|@{~~}c@{~~}||@{~~}c@{~~}|@{~~}c@{~~}|@{~~}c@{~~}|@{~~}c@{~~}||c|c|} \hline
    coefficient & $c_3$ & $c_2$ & $c_1$ & \hspace{-0.8em} $-c_0$ & cost & stability \\ \hline
    degeneracy & $8$ & $12$ & $6$ & $1$ & $(\times,+)$ & $(\alpha > \ldots)$ \\ \hline\hline
    $\Big.$ \textbf{standard} $\Big.$ & $0$ & $0$ & $1$ & $6$ & $1,6$ & $\sqrt{3}$ \\ \hline
    $\Big.$ \textbf{isotropic A} $\Big.$ & $0$ & $\frac{1}{6}$ & $\frac{1}{3}$ & $4$ & $3,18$ & $\sqrt{2}$ \\ \hline
    $\Big.$ \textbf{isotropic B} $\Big.$ & $\frac{1}{12}$ & $0$ & $\frac{2}{3}$ & $\frac{14}{3}$ & $3,14$ & $\sqrt{21}/3$ \\ \hline
    $\Big.$ \textbf{isotropic C} $\Big.$ & $\frac{1}{30}$ & $\frac{1}{10}$ & $\frac{7}{15}$ & $\frac{64}{15}$ & $4,26$ & $8/\sqrt{30}$ \\ \hline
  \end{tabular}
  \caption{Summary of spatial discretization schemes.}
  \label{tab:disc}
\end{table}

\noindent
Discretization of spatial differential operators
\begin{equation}
  \Delta X = \frac{D[X]}{(dx)^2}, \hspace{1em}
  (\vec{\nabla} X)^2 = \frac{G[X]}{(dx)^2}
\end{equation}
allows more freedom. Direct generalization of the second derivative
discretization in equation (\ref{eq:disc}) to three spatial dimensions leads
to the often-used second-order accurate expression for Laplacian using six
nearest neighbours of a point. However, this is not the only (or the best)
choice. Truncation error for this scheme depends on direction, introducing
anisotropic artifacts in the field evolution at short length scales. One can
mitigate this unpleasant fact by increasing resolution, but smarter
discretization scheme is a better solution. By using all 26 neighbours in a $3
\times 3 \times 3$ cube around a point, one can derive a family of
discretizations of Laplacian operator which is second-order accurate and {\em
fourth-order isotropic} \cite{Patra:2005}. For discretization to be isotropic,
the coefficients in a linear combination $D[X]$ approximating the Laplacian
operator should only depend on the distance from the central point
\begin{equation}\label{eq:D}
  D[X] \equiv \underbrace{\sum\limits_{x-1}^{x+1}\sum\limits_{y-1}^{y+1}\sum\limits_{z-1}^{z+1}}\limits_\alpha
    c_{\text{d}(\alpha)} X_\alpha,
\end{equation}
as illustrated in Figure~\ref{fig:disc}. The values of coefficients for
possible discretizations of the Laplacian operator are summarized in
Table~\ref{tab:disc}, along with their computational cost and stability
properties. Isotropic discretizations $A$ and $B$ offer reduced computational
cost, while isotropic discretization $C$ has the best accuracy and stability.
As multiplications are cheap and additions are essentially free on modern
CPUs, there is no reason not to use the best discretization scheme available.
Thus, DEFROST is configured to use the isotropic discretization $C$ by
default. That can be changed by uncommenting the other coefficient definitions
in the source code, although profiling shows little gain in doing so for large
grids (for which performance of DEFROST solver is apparently memory-bandwidth
dominated).

To calculate the energy density (\ref{eq:rho}), we need to discretize the
square of spatial gradients as well. It is very important for discretized
energy to be conserved by discretized equations of motion. Otherwise, it will
leak off the grid in the course of a long simulation, affecting overall
accuracy or even giving incorrect results (for equation of state, for
example). Simply squaring the first spatial derivative discretized like in
equation (\ref{eq:disc}), although second-order accurate, is {\em not}
conservative. Using discretized action approach \cite{Pen:1993nx}, one can
show that the conservative second-order accurate and fourth-order isotropic
discretization of the gradient-square operator is
\begin{equation}\label{eq:G}
  G[X] \equiv \frac{1}{2} \underbrace{\sum\limits_{x-1}^{x+1}\sum\limits_{y-1}^{y+1}\sum\limits_{z-1}^{z+1}}\limits_\alpha c_{\text{d}(\alpha)} (X_\alpha-X_0)^2,
\end{equation}
where coefficients $c_i$ are the same as in Laplacian (\ref{eq:D}). Evaluating
this expression requires 30 multiplies per point for discretization $C$, which
is significantly more expensive than computing the Laplacian. However, this
does not place much of a burden on the total runtime. As we mentioned before,
we eliminated gradient-square terms from evolution equations, so they only
need to be calculated for output, which happens much less often.

Putting everything together, we end up with the following evolution scheme:
the discretized field values are advanced to the next time slice using
\begin{equation}\label{eq:step}
  \phi^i_{\text{up}} =
  \frac{2 + D/(\alpha a)^2 - M_i^2 dt^2}{1+\frac{3}{2}\,Hdt}\, \phi^i_{\text{hr}} -
  \frac{1-\frac{3}{2}\,Hdt}{1+\frac{3}{2}\,Hdt}\, \phi^i_{\text{dn}},
\end{equation}
where $D$ is the discretized Laplace operator (\ref{eq:D}) and $\alpha \equiv
dx/dt$. The time step $dt$ has to be small enough both to satisfy Courant stability
condition ($\alpha > \text{const}$ listed in Table~\ref{tab:disc}) and to
resolve the period of the fastest oscillating field. All the coefficients
in expression (\ref{eq:step}) except for the effective mass of the $i$-th
field
\begin{equation}\label{eq:M}
  M_i^2 \equiv \frac{1}{\phi^i} \frac{\partial V}{\partial \phi^i}
\end{equation}
are constant on the grid and the same for all fields, and thus can be
pre-computed outside the evolution loop. Inside the same loop, kinetic and
potential energy of the fields $\phi^i$ are accumulated to evaluate
\begin{equation}
  \langle\rho + 3p\rangle_{\text{hr}} = \frac{1}{n^3} \sum_{x,y,z} \left[
    \sum_i \frac{\left(\phi^i_{\text{up}} - \phi^i_{\text{dn}}\right)^2}{2\, dt^2} - 2 V(\phi^i) \right],
\end{equation}
which is used to advance the Hubble length
\begin{equation}\label{eq:step:L}
  L_{\text{up}} = L_{\text{dn}} +
    \left[ 1 + \frac{\bar{L}_{\text{hr}}^2}{6}\, \langle\rho + 3p\rangle_{\text{hr}} \right] 2\, dt.
\end{equation}
To avoid weak numerical instability associated with even-odd slice decoupling,
we use $\bar{L}_{\text{hr}} = (L_{\text{up}}+L_{\text{dn}})/2$ instead of
$L_{\text{hr}}$ in the above equation, which then can be solved for
$L_{\text{up}}$ either as an exact quadratic or iteratively (we use the latter
in the code).

Equations (\ref{eq:step}) and (\ref{eq:step:L}) provide a complete recipe how
to advance the field variables to the next time step. Once in a while (at
user's request) we would also like to calculate, analyze, and output for
visualization some auxiliary variables like energy density $\rho$ and
gravitational potential $\Psi$. Energy density (\ref{eq:rho}) and pressure
(\ref{eq:p}) are easy to calculate once we know the field values $\phi^i$ and
their gradients (\ref{eq:G}). Finding the gravitational potential is a little
less trivial, as we need to solve the Laplace equation (\ref{eq:Psi}) with
periodic boundary conditions. The fastest way to do it is to use a fast
Fourier transform (FFT). Applying FFT to the discretized Laplacian operator
(\ref{eq:D}), we end up with an algebraic equation for $\Psi$ in Fourier
domain
\begin{equation}
  \Psi_k = \frac{\rho_k\, a^2\, dx^2}{2\, P(\cos\frac{2\pi}{n} k_x, \cos\frac{2\pi}{n} k_y, \cos\frac{2\pi}{n} k_z)}
\end{equation}
where $k \in [0,n)$ is the integer Fourier mode wave-number
and polynomial $P(i,j,k)$ follows from discretization (\ref{eq:D})
\begin{equation}
  P(i,j,k) = \tilde{c}_0 + \tilde{c}_1 (i+j+k) + \tilde{c}_2 (ij+ik+jk) + \tilde{c}_3\, ijk,
\end{equation}
with $\tilde{c}_k = 2^k c_k$. Here $c_k$ are once again the coefficients of
discretization (\ref{eq:D}) with values listed in Table~\ref{tab:disc}.

\begin{figure}
  \centerline{\epsfig{file=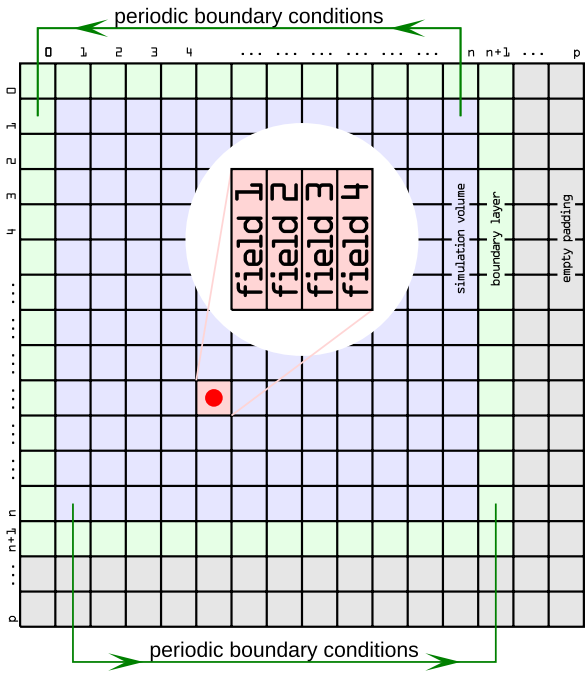, width=220pt}}
  \caption{
    Memory layout of the spatial grid and field samples.
  }
  \label{fig:layout}
\end{figure}

DEFROST implements the evolution scheme (\ref{eq:step}) in Fortran 90, fully
taking advantage of capabilities of modern hardware and compilers by using both
automatic vectorization (over field variables) and automatic parallelization (of
the evolution loop). During code development, profiling showed that physical
memory layout has an unexpectedly large impact on performance, which became
apparent as the solver loop got optimized. Let us briefly discuss the storage
model which was adopted after some investigation. The fields $\phi^i$ are
sampled and stored in a large multi-dimensional array \texttt{smp(N,0:p,0:p,0:p,3)}.
The first index (minor in Fortran index ordering) enumerates the $N$ fields.
The next three indices enumerate the three dimensions of the spatial grid,
padded to $(p+1)^3$ elements for reasons discussed below. The last index
(major in Fortran index ordering) enumerates the three time slices used in the
evolution scheme. Rather than copying large amounts of data around, the
allocation of indices for \texttt{dn}, \texttt{hr}, and \texttt{up} slices
cycles every time step.

The layout of a single time slice is shown in Figure~\ref{fig:layout} (with
one spatial dimension suppressed for clarity). The $n \times n \times n$
simulation volume is surrounded by a single cell wide boundary layer,
introduced to implement the boundary conditions without conditional logic
inside the evolution loop. It also allows for an easy transition to parallel
cluster implementation using MPI, as required buffers are already in place.
Somewhat counter-intuitively, padding the grid by a few extra empty cells can
significantly improve the runtime for large grids. The root cause for this
phenomenon probably lies in some interaction between memory access pattern and
hardware memory cache algorithm. To get the best FFT performance, the grid
size $n$ is usually taken to be a power of two. While evaluating
(\ref{eq:step}) on an un-padded array, memory would be accessed with a power
of two stride, which could conceivably interfere with caching and prefetching
done by the memory subsystem. Whatever the cause, padding the array so that
its size $p+1$ is prime (or a product of a few large primes) can improve the
runtime, so the user is advised to experiment.

Finally, a few words should be said about statistical estimators used to
analyze the simulation data. The power spectral density (PSD) estimator is a
fairly conventional one implemented using FFT. It employs anti-aliasing with
fourth-order polynomial kernel when folding the spectrum into wave-number bins
to reduce sampling noise. The implementation of probability density function
(PDF) estimator is less conventional, and does not use histogram binning at all.
Instead, PDF is derived from cumulative density function (CDF) which is
obtained by partially sorting the data cube into $n$ quantile brackets.
Although more expensive and harder to parallelize, this approach is more
robust and offers uniform sampling noise across the distribution.

\section{Initial Conditions}\label{sec:ic}

At the end of the inflation, most of the energy is still stored in the
inflaton, and all the fields are homogeneous except for small quantum
fluctuations. But the presence of these quantum fluctuations in the fields is
essential to trigger the dynamical instability leading to preheating.

Initial conditions in the homogeneous field components depend on the model of
inflation and are treated as an external input by DEFROST. They are
straightforward to obtain by following the expanding Friedman-Robertson-Walker
solution during inflation either analytically (if possible), or using any of
the available numerical ordinary differential equation integrators. As
inflationary trajectory is an attractor \cite{Belinsky:1985zd}, it is easy to find and no particular
care is needed on where to start tracing it. The full three-dimensional
simulation of preheating should take over at some time near the end of the
inflation, but as there are other factors at play (such as limited spatial
dynamic range available to simulation), the exact moment is best decided on
case by case basis.

\begin{figure}
  \centerline{\epsfig{file=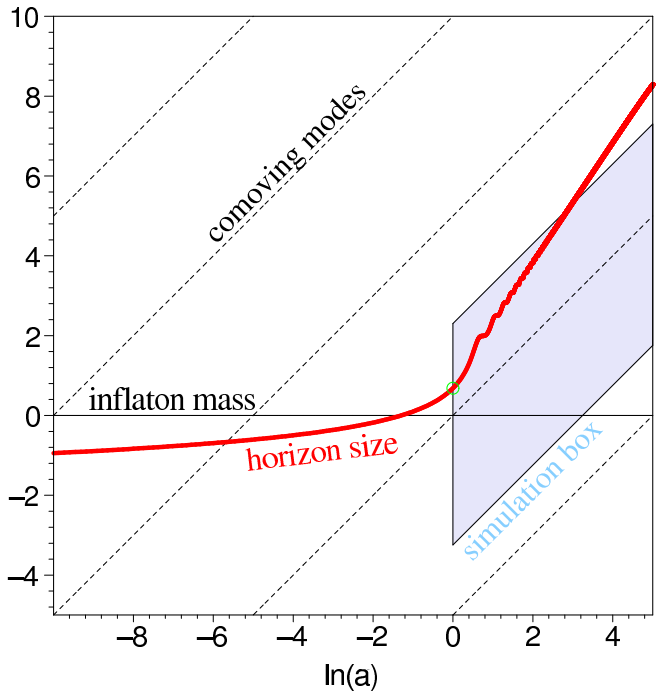, width=246pt}}
  \caption{
    Expansion history of the universe in chaotic inflation. Numerical
    simulation starts at the end of inflation ($\epsilon=-1$). The size of
    comoving simulation box is selected to cover the scales corresponding to
    horizon size (red), inflaton mass (thin black line), and the wave modes in
    unstable band (not shown).
  }
  \label{fig:exp}
\end{figure}

To make a concrete example, Figure~\ref{fig:exp} shows expansion history of a
typical chaotic inflation model with massive inflaton $V(\phi)=m^2\phi^2/2$
(discussed in more detail in the next section). As the Universe is inflating,
its horizon size $L \equiv 1/H$ stays relatively constant, while the physical
wavelength of comoving modes grows with the scale factor $a$. The modes which
were originally inside the horizon expand and leave the horizon during
inflation. Eventually, inflation ends and the horizon size starts growing
faster than the scale factor (for instance, $L\propto a^{3/2}$ during matter
domination), at which point the modes begin to re-enter the horizon. The
moment in time when comoving modes stop leaving the horizon and begin re-entry
can be taken as the end of the inflation. This happens when
\begin{equation}
  \frac{d\ln L}{d\ln a} = 1,
\end{equation}
or in terms of a slow roll parameter
\begin{equation}\label{eq:end}
  \epsilon \equiv \frac{\dot{H}}{H^2} = -1.
\end{equation}
For simulations presented in this paper, we start exactly at the moment when
inflation ends (\ref{eq:end}), and select the comoving box size of $\ell=10/m$
to cover all the length scales of interest, as illustrated in Figure~\ref{fig:exp}.

As we already mentioned, it is crucial to include quantum field fluctuations
in the mostly-homogeneous initial conditions for the preheating instability to
develop. The spectra of quantum field fluctuations are determined by effective
masses of the fields involved. Let us briefly recount the standard derivation
\cite{Linde:2005ht} to establish notation. Canonically normalized massive
field $\varphi$ with Lagrangian
\begin{equation}
  {\cal L} = \frac{1}{2} (\nabla\varphi)^2 - \frac{1}{2} m^2 \varphi^2
\end{equation}
is quantized in flat Minkowski spacetime by promoting the field value $\varphi$ and field momentum
\begin{equation}
  \pi \equiv \frac{\partial\cal L}{\partial\dot\varphi} = \dot{\varphi}
\end{equation}
to quantum operators $\hat{\varphi}$ and $\hat{\pi}$ obeying canonical commutation relation
\begin{equation}\label{eq:canonical}
  [\hat{\varphi}(\vec{x}), \hat{\pi}(\vec{x}')] = i\, \delta(\vec{x}-\vec{x}').
\end{equation}
In flat spacetime, the field operator can be represented using Fourier mode decomposition as
\begin{equation}\label{eq:op:phi}
  \hat{\varphi}(x^\mu) = \frac{1}{(2\pi)^{\frac{3}{2}}} \int \frac{d^3k}{\sqrt{2\omega}}
    \left[e^{i k_\mu x^\mu} \hat{a}^+_k + e^{-i k_\mu x^\mu} \hat{a}^-_k \right],
\end{equation}
where the mode creation-annihilation operators $\hat{a}^\pm$ obey
\begin{equation}\label{eq:op:a}
  [\hat{a}^-_k, \hat{a}^+_q] = \delta(\vec{k}-\vec{q}),
\end{equation}
which directly follows from canonical commutation relation (\ref{eq:canonical}).
The mode frequency $\omega$ of the massive field is related to its wavevector
$\vec{k}$ by a simple dispersion relation
\begin{equation}
  \omega^2 = m^2 + k^2.
\end{equation}
Using mode decomposition (\ref{eq:op:phi}) and commutation relation
(\ref{eq:op:a}), it is easy to show that the two-point correlation functions
of the field value and momentum are
\begin{eqnarray}
  \label{eq:2pt:phi}
  \langle\hat{\varphi}(\vec{x})\hat{\varphi}(\vec{x}')\rangle &=&
    \frac{1}{(2\pi)^3} \int \frac{d^3 k}{2\omega}\, e^{i\vec{k}(\vec{x}-\vec{x}')}, \\
  \label{eq:2pt:pi}
  \langle\hat{\pi}(\vec{x})\hat{\pi}(\vec{x}')\rangle &=&
    \frac{1}{(2\pi)^3} \int \frac{d^3 k}{2}\, \omega\, e^{i\vec{k}(\vec{x}-\vec{x}')}.
\end{eqnarray}
The spectrum of the field $\varphi$ fluctuations is simply $1/(2\omega)$.

One can repeat this procedure in the background of expanding homogeneous
universe. In general, the time dependence of the modes will be different, but
it turns out that quantum fluctuations of massive fields in de Sitter
spacetime can be approximated quite well with above expressions if one simply
replaces the field mass $m$ with an effective mass
\begin{equation}
  m_{\text{eff}}^2 = m^2 - \frac{9}{4}\, H^2.
\end{equation}
We will use this approximation in DEFROST. In addition, we will follow the
established practice of treating quantum operators as Gaussian random
variables. This is an assumption, but not totally unjustified one. Quantum
modes essentially behave classically after they leave the horizon
\cite{Polarski:1995jg}. Even the sub-horizon modes (which are initially
quantum) get large occupation numbers once the preheating instability kicks
in, and can be treated classically \cite{Khlebnikov:1996mc}. Thus, we will
initialize the field fluctuations as a Gaussian random field
\begin{equation}\label{eq:random}
  \hat{\varphi}(\vec{x},t) = \frac{1}{(2\pi)^{\frac{3}{2}}} \int \frac{d^3k}{\sqrt{2\omega}}
    e^{i\vec{k}\vec{x}} \left[\hat{b}_k \cos\omega t + \hat{c}_k \sin\omega t\right],
\end{equation}
where the complex random operators $\hat{b}_k$ and $\hat{c}_k$ obey
\begin{equation}
  \langle \hat{b}_k \hat{b}_{k'}^*\rangle =
  \langle \hat{c}_k \hat{c}_{k'}^*\rangle = \delta(\vec{k}-\vec{k}')
\end{equation}
to reproduce the two-point correlation functions (\ref{eq:2pt:phi}, \ref{eq:2pt:pi}).

A lot of effort has gone into making the realization of random field initial
conditions in DEFROST as statistically accurate as possible. The
straightforward and often used way to generate a Gaussian random field on a
discrete grid is to directly discretize equation (\ref{eq:random}) in Fourier
space, assigning $k$-th mode a random Gaussian number with amplitude
$1/\sqrt{2\omega}$. Although simple, this procedure is spoiled by the finite
grid size effects, and does not reproduce correct two-point correlation
functions in the real space \cite{Pen:1997up}. One ends up with a substantial
lack of power on the scales comparable with the box size, which is not
surprising if one considers that only a few long-wavelength modes ``fit'' into
the box, and naive discretization ignores all the power in the infrared part
of the spectrum which should have been properly aliased into those few
low-$k$ modes.

A wealth of the literature is dedicated to the subject of generating Gaussian
random fields of a given spectrum, particularly in the context of the $N$-body
simulations of the large scale structure \cite{Fournier:1982,
Efstathiou:1985re, Hoffman:1991, Hoffman:1992, Salmon:1996, Pen:1997up,
Bertschinger:2001ng, Plaszczynski:2005yp}. We draw on that experience, and in
DEFROST for random field generator we adopt a method described in
\cite{Salmon:1996,Pen:1997up}. Gaussian random field (\ref{eq:random}) with
$1/(2\omega)$ spectrum is realized by convolving white noise with a
spherically symmetric kernel function
\begin{eqnarray}
  \xi(r) &=& \frac{1}{(2\pi)^{\frac{3}{2}}} \int \frac{d^3k}{\sqrt{2\omega}} e^{i\vec{k}\vec{x}}\\
         &=& \frac{1}{\sqrt{\pi}} \int \frac{k^2 dk}{(k^2 + m_{\text{eff}}^2)^\frac{1}{4}} \frac{\sin kr}{kr}.\nonumber
\end{eqnarray}
The kernel function $\xi(r)$ can be evaluated analytically in terms of Bessel functions
\begin{equation}
  \xi(r) = \frac{2^{\frac{3}{4}} m^{\frac{1}{4}}}{4\pi\, r^{\frac{9}{4}}}\, \Gamma\left({\textstyle\frac{3}{4}}\right)
    \left[ K_{\frac{1}{4}}(mr) + 2 mr K_{\frac{3}{4}}(mr) \right],
\end{equation}
and has a power law ultraviolet divergence $\xi(r) \propto r^{-\frac{5}{2}}$
in the limit $r \rightarrow 0$. For the purposes of discretization on a finite
grid, we have to regularize this divergence, which we do by introducing a
Gaussian cut-off at some scale $q$ below the Nyquist frequency
\begin{equation}
  \xi(r) = \frac{1}{\sqrt{\pi}} \int \frac{k^2 dk}{(k^2 + m_{\text{eff}}^2)^\frac{1}{4}} \frac{\sin kr}{kr}\, \exp\left[-\frac{k^2}{q^2}\right].
\end{equation}
The regularized kernel does not have a nice expression in terms of elementary
functions, but is easy to evaluate numerically, of course. We use a
one-dimensional discrete sine transform (DST) on a substantially larger grid
to calculate it (simply because DST is already provided by FFTW libraries we
use), but other methods like quadrature integrators could be used as well.
Once the spherically-symmetric kernel $\xi(r)$ is evaluated, it is sampled on
the the three-dimensional grid in real space, and the random field is initialized as a
convolution
\begin{equation}
  \hat{\varphi}(\vec{x},0) = \frac{1}{(2\pi)^3} \iint d^3k\, d^3x'\, \hat{b}_k \xi(\vec{x}') e^{i\vec{k}(\vec{x}-\vec{x}')}.
\end{equation}
The convolution is implemented using discrete FFT as
\begin{equation}
  \varphi(\vec{x},0) = \sum\limits_{\vec{k}}\sum\limits_{\vec{x}'} \frac{B_k \xi(\vec{x}') e^{i\vec{k}(\vec{x}-\vec{x}')}}{2^{\frac{1}{2}} (dk)^{\frac{3}{2}} n^3},
\end{equation}
where $n$ is the grid size, $dk = 2\pi/\ell$ is the spacing of discrete
wavemodes, and $B_k$ is a complex Gaussian random number generated using
Box-Muller transformation
\begin{equation}\label{eq:box-muller}
  B_k = \sqrt{-2\ln U_1}\, e^{2\pi i U_2}
\end{equation}
from two real random numbers $U_1$ and $U_2$ uniformly distributed on a unit
interval. The implementation of initial velocity generator is entirely
analogous (and handled by the same procedure), and is not worth repeating
here.

Finally, I have to point out that as of current writing, there is a bug in the
random number generator implementation in LATTICEEASY. The formula
(\ref{eq:box-muller}) is modified there to produce {\em two} complex numbers using
only {\em three} uniformly distributed real numbers. This results in
correlated random numbers with quite non-Gaussian distribution. Fortunately,
the results reported so far do not seem to be too much affected by this problem, but
as non-gaussianity studies are becoming more prominent in the modern cosmology
literature, some care must be taken in proper implementation of random number
generators.

\section{Chaotic Inflation and Broad Parametric Resonance}\label{sec:chaotic}

So far, the discussion was rather general, as DEFROST is designed to be easily
adoptable to study arbitrary models of preheating. This Section describes
preheating model we selected for first simulations with DEFROST: chaotic
inflation ending via broad parametric resonance \cite{Kofman:1994rk}. The
development of linear instability in this model can be largely understood
analytically \cite{Kofman:1997yn}, and its non-linear dynamics has been widely
studied numerically as well \cite{Podolsky:2005bw,Felder:2006cc}. For all its
simplicity, this model has very rich dynamics, and still holds surprises. Our
very first simulations uncover new aspects of the evolution dynamics in this
model, which are reported in the next Section.

\begin{figure}
  \centerline{\epsfig{file=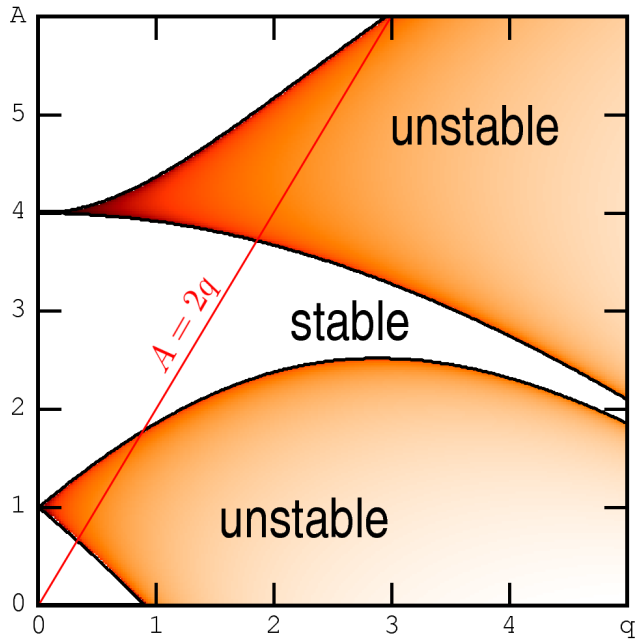, width=246pt}}
  \caption{
    Stability bands of Mathieu equation.
  }
  \label{fig:mathieu}
\end{figure}

In a minimal form, the model consists of two scalar fields: the massive
inflaton $\phi$ and the massless decay product $\psi$ interacting via
potential
\begin{equation}\label{eq:model}
  V(\phi,\psi) = \frac{1}{2}\, m^2 \phi^2 + \frac{1}{2}\, g^2 \phi^2 \psi^2.
\end{equation}
During inflation, the value of the inflaton $\phi$ is large, the field is
over-damped by the large Hubble friction, and slowly rolls down its potential.
As it reaches the value of around one in Planck units, the damping dips below
critical, and homogeneous inflaton starts oscillating with decreasing
amplitude
\begin{equation}
  \phi(t) \approx \Phi(t) \sin mt, \hspace{1em}
  \Phi(t) = \frac{\Phi_0}{a^\frac{3}{2}} = \sqrt{\frac{2}{3}}\, \frac{2}{mt}.
\end{equation}
Decay field $\psi$ is coupled to inflaton, and feels its oscillations through
modulation of the effective mass; equation of motion for the Fourier mode
$\psi_k$ with wavector $\vec{k}$ is
\begin{equation}\label{eq:psi_k}
  \ddot{\psi}_k + 3H\dot{\psi}_k + \left(\frac{k^2}{a^2} + g^2\Phi^2\sin^2mt\right) \psi_k = 0.
\end{equation}
If the coupling $g$ is large enough, periodic modulation of the field mass
leads to strong instability via parametric resonance. This can be understood
analytically by applying general theory of differential equations with
periodic coefficients \cite{Kofman:1997yn}. If one ignores the expansion and
the Hubble drag term in equation (\ref{eq:psi_k}), evolution for the Fourier
mode $\psi_k$ is given by Mathieu equation
\begin{equation}\label{eq:mathieu}
  \frac{d^2\psi_k}{d\eta^2} + (A - 2q \cos 2\eta) \psi_k = 0,
\end{equation}
where we have introduced dimensionless parameters
\begin{equation}\label{eq:params}
  A = 2q + \frac{k^2}{m^2 a^2}, \hspace{1em}
  q = \frac{g^2\Phi^2}{4 m^2},
\end{equation}
and time variable $\eta \equiv mt$ to bring the equation into canonical form.
According to Flouqet's theorem, a general solution of Mathieu equation
(\ref{eq:mathieu}) is of the form $e^{\mu \eta} P(\eta)$, where $P(\eta)$ is a
periodic function with period $\pi$. Floquet exponent $\mu$ depends on
parameters $A$ and $q$, and there is an elegant way to calculate its value
\cite{Bateman:1955}, which we (somewhat reluctantly) will omit here, and just
quote the final result. Figure~\ref{fig:mathieu} shows the dependence of
$\text{Re}\, \mu$ on parameters $A$ and $q$ as a density plot. For certain
parameter values Floquet exponent $\mu$ has positive real part, leading to
exponential instability of the solution; these unstable bands are marked on
Figure~\ref{fig:mathieu}. The value of $A$ for our problem (\ref{eq:params})
is restricted to lie above the red line $A=2q$ in Figure~\ref{fig:mathieu},
which corresponds to the homogeneous mode with $k=0$. For sufficiently large
coupling $q \gg 1$, large portion of available phase space volume is unstable,
leading to fast development of instability. This regime is known as broad
parametric resonance. The instability grows on a timescale comparable to $1/m$
(as Floquet exponent values are around $\text{Re}\, \mu \lesssim 1/3$), and
manifests itself after a few dozen of oscillations of the inflaton, which is
very fast in cosmological terms. In the next Section, we describe the
non-linear field evolution after this instability develops.

\section{Numerical Results}\label{sec:results}

Before we present our simulation results, a few words should be said about the
units used throughout the code. As it is clear from the action
(\ref{eq:action}), we prefer to work with dimensionless scalar fields, rather
than canonically normalized ones. In this convention, one can simply think of
the values of the scalar fields as measured in units of Planck mass
$m_{\text{pl}}$. Note that we use reduced Planck mass $m_{\text{pl}} = (8\pi
G)^{-1/2}$ rather than $M_{\text{pl}} = G^{-1/2}$. The only other scale in the
model (\ref{eq:model}) is the inflaton mass $m$. The coupling constant $g$
(which has dimension of mass with our scalar field normalization), the
frequency and wavevectors of the unstable modes, and everything else can be
referred to it. It is convenient to scale all the quantities (except for field
values) by the appropriate power of $m$, making them dimensionless and
suitable for numerical analysis. Thus we set $m=1$ in the code.

We simulate the preheating model (\ref{eq:model}) with inflaton mass $m = 5
\cdot 10^{-6}\, m_{\text{pl}}$, the value of coupling constant $g=100\,m$ in a
comoving box of size $\ell=10/m$ (the values chosen correspond to the ones
used in Ref.~\cite{Felder:2006cc}). The grid size is taken to be $256^3$,
while the time step $dt = 2^{-10}/m$ has to be reduced substantially below the
Courant limit to resolve oscillations of the field $\psi$ (which is initially
hundred times heavier than inflaton $\phi$). The simulation is started at the
end of the inflation (\ref{eq:end}), when the value of the inflaton is $\phi
\approx 1.009343$ and the Hubble constant is $H \approx 0.50467\,m$. The
simulation box is initially about five Hubble lengths across. We let the code
run until $t=256/m$, which corresponds to $2^{18}$ time steps.

\begin{figure*}
  \begin{center}
  \begin{tabular}{cc}
    \epsfig{file=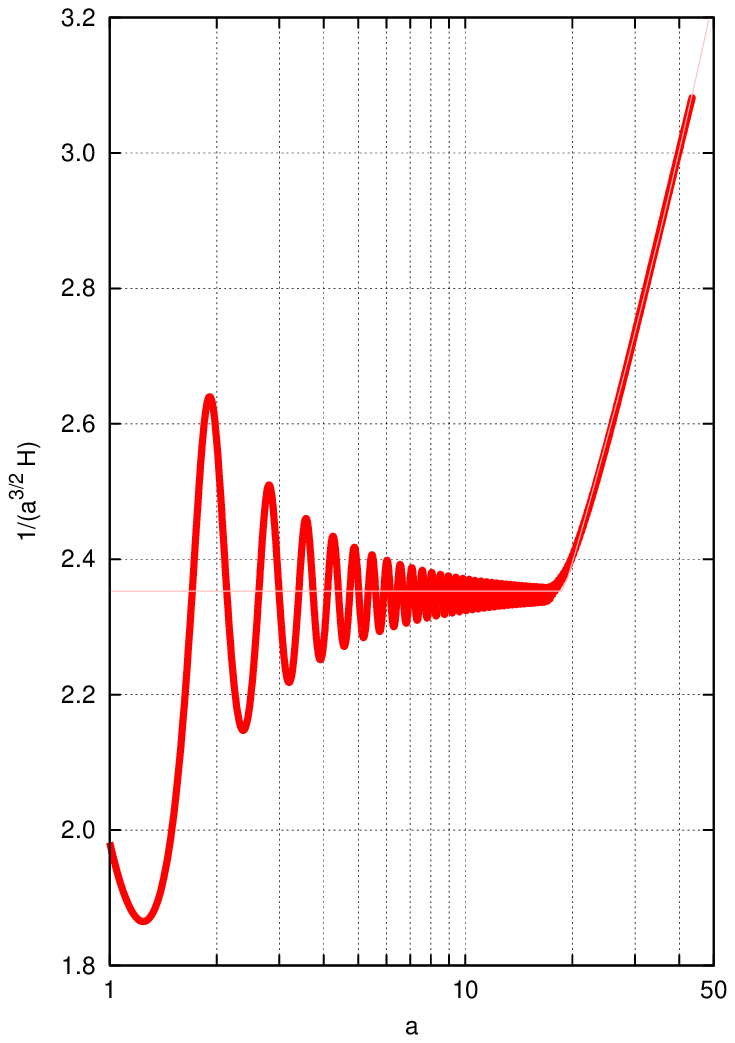, width=246pt} &
    \epsfig{file=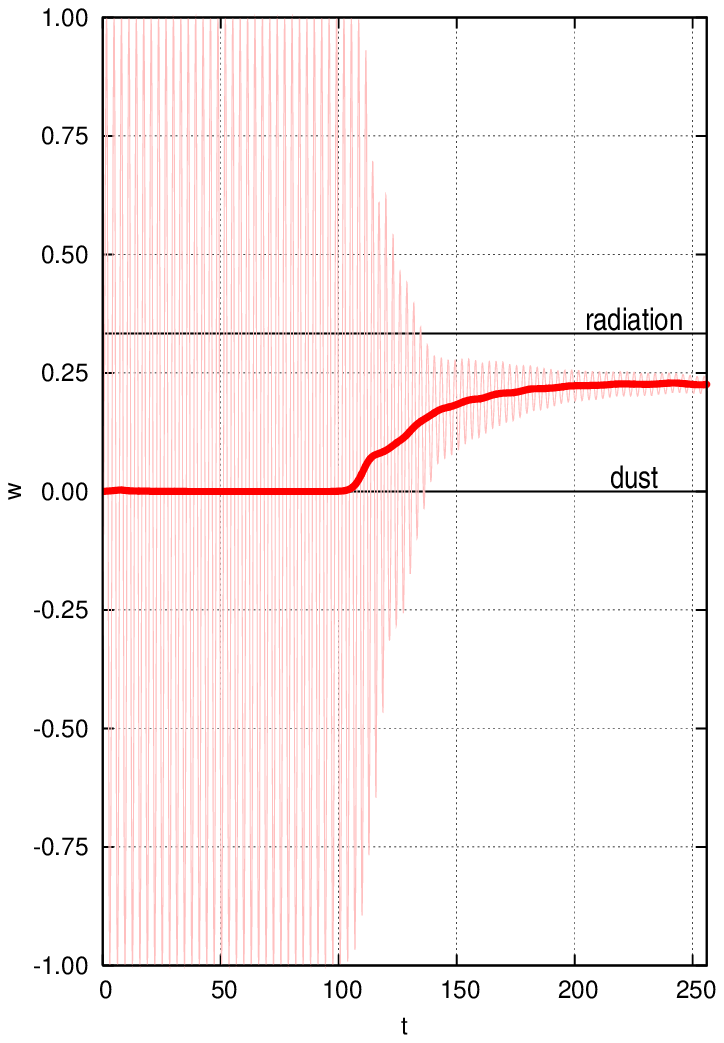, width=246pt}
  \end{tabular}
  \end{center}
  \vspace{-12pt}
  \caption{
    Expansion history (left) and average equation of state (right) during preheating.
  }
  \label{fig:eos}
\end{figure*}

To get things going, we first reproduce the previously reported results
\cite{Podolsky:2005bw} on expansion history of the preheating model
(\ref{eq:model}). The left panel of Figure~\ref{fig:eos} shows evolution of
horizon size during expansion, with $a^{3/2}$ growth corresponding to
matter-dominated expansion scaled out. The expansion history has a sharp break
as instability develops, and energy gets deposited into relativistic
inhomogeneous modes from a homogeneous oscillating inflaton (which behaves as
a pressureless dust). This transition can be seen in terms of an effective
equation of state parameter $w \equiv \langle p \rangle/\langle \rho \rangle$,
the value of which is plotted in the right panel of Figure~\ref{fig:eos}. It
undergoes large amplitude oscillations (shown by thin pale line in
Figure~\ref{fig:eos}), but when averaged over a few periods (with a Kaiser
window function), the underlying behavior is uncovered. The averaged equation
of state (shown by thick red line) switches over from dust-like equation of
state $w=0$ to a value slightly less than a quarter \cite{Podolsky:2005bw}, corresponding to a fairly
relativistic fluid. (If evolved further, the residual homogeneous component in
the inflaton will eventually come to dominate the evolution again, slowly
lowering equation of state toward $w=0$ in the process.)

\begin{figure}[b]
  \centerline{\epsfig{file=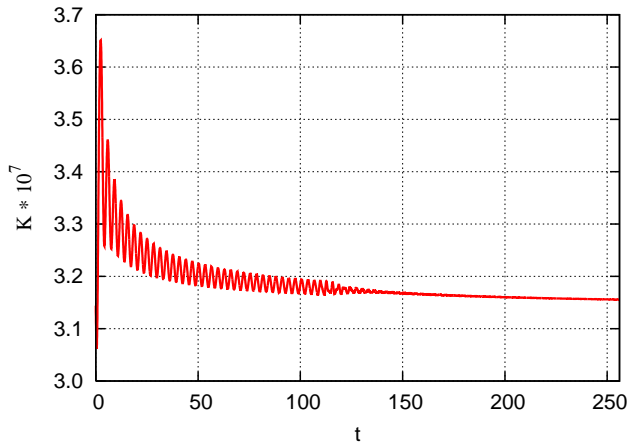, width=246pt}}
  \vspace{-6pt}
  \caption{
    Residual curvature $K$ testing constraint violation.
  }
  \label{fig:flat}
\end{figure}

While we recover the results obtained by simulations with LATTICEEASY, the
accuracy of the integrator used in DEFROST is significantly higher. The
performance of integration scheme for expansion factor (\ref{eq:step:L}) is
illustrated in Figure~\ref{fig:flat}, which shows residual curvature
\begin{equation}
  K = a^2 \left(\frac{\langle\rho\rangle}{3} - H^2\right),
\end{equation}
which should be zero in the flat model we are evolving. As you can see, the
constraint equation is satisfied to $10^{-7}$ level (which is exceptionally
good for a second order scheme with $10^{-3}$ timestep), and the error does not
accumulate with time. In fact, this error is mostly due to the fact that we
neglected second order corrections to density from initial field fluctuations.

\begin{figure*}
  \begin{center}
  \begin{tabular}{cc}
    \epsfig{file=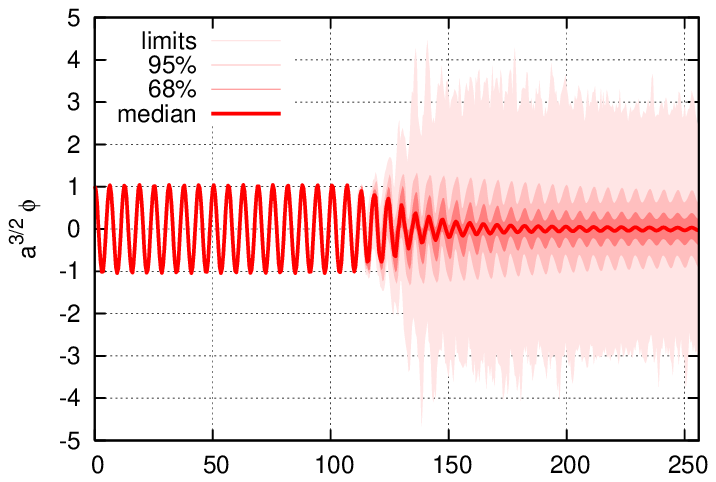, width=246pt} &
    \epsfig{file=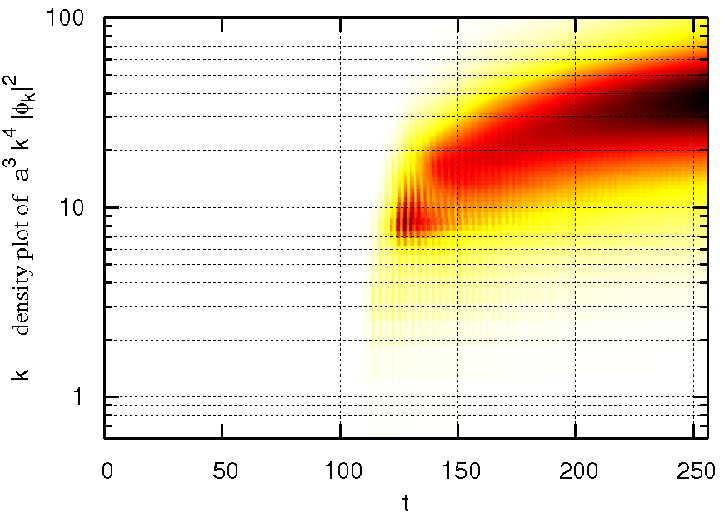, width=246pt}\vspace{-21pt}\\
    \epsfig{file=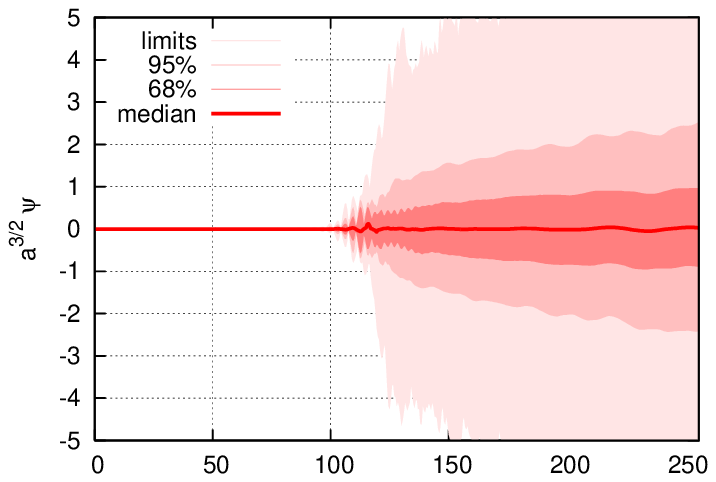, width=246pt} &
    \epsfig{file=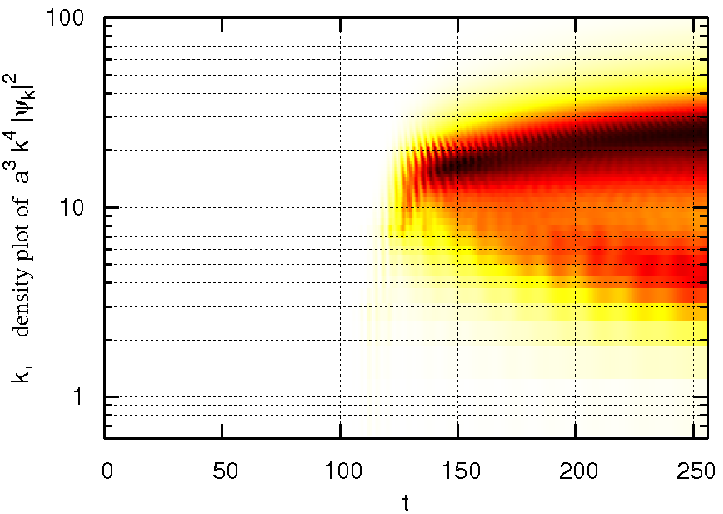, width=246pt}\vspace{-21pt}\\
    \epsfig{file=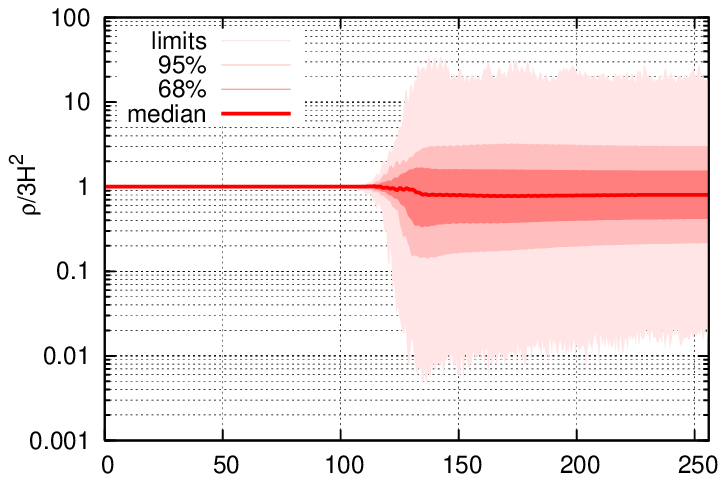, width=246pt} &
    \epsfig{file=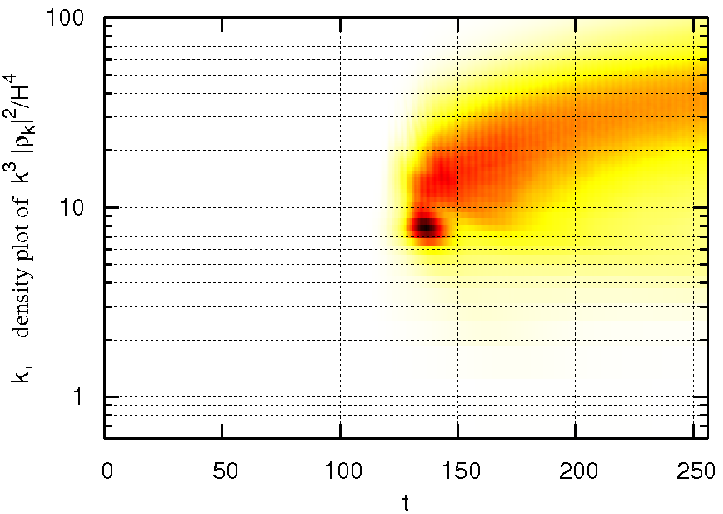, width=246pt}\vspace{-21pt}\\
    \epsfig{file=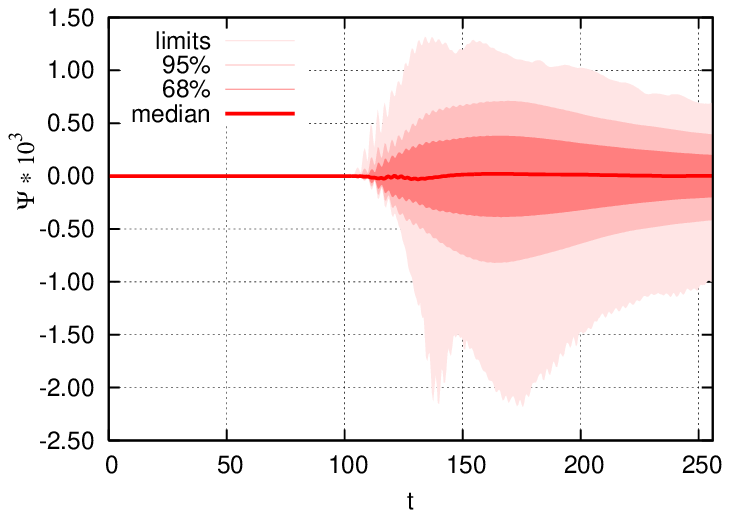, width=246pt} &
    \epsfig{file=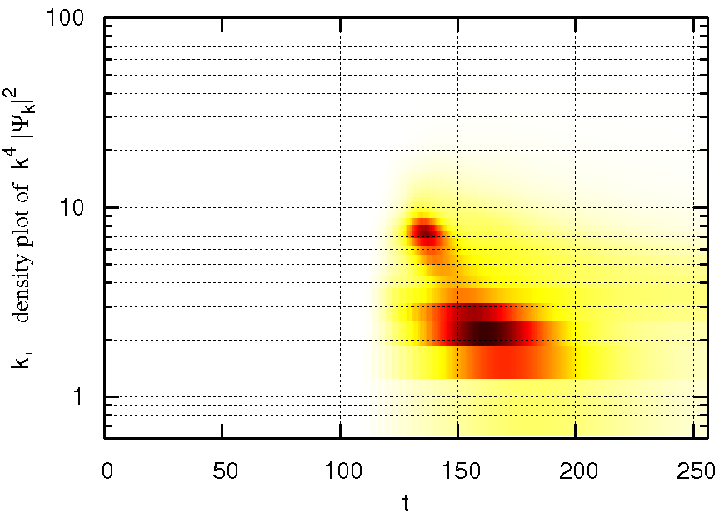, width=246pt}\vspace{-16pt}\\
  \end{tabular}
  \end{center}
  \caption{
    Evolution of field value distributions (left) and spectra (right) of
    inflaton $\phi$ (top row), decay field $\psi$ (second row), total density
    $\rho$ (third row), and gravitational potential $\Psi$ (bottom row). Onset
    of instability and characteristic size of the structure formed is clearly
    visible on the plots.
  }
  \label{fig:stats}
\end{figure*}

Having made sure that our code reproduces previously reported results, and
after numerous checks of code integrity and accuracy, we move on to
investigation of the field dynamics during preheating.

Evolution of field distributions and spectra for inflaton $\phi$, decay field
$\psi$, total density $\rho$ and gravitational potential $\Psi$ are presented
in Figure~\ref{fig:stats}. Left panel shows evolution of the median value
(thick red line), along with 68\% and 95\% percentile brackets around it
(which would correspond to $1\sigma$ and $2\sigma$ contours for a Gaussian
distribution) shown by shaded outlines. The contour with the lightest shading
spans the extremal values inside the simulation box, which serves to
illustrate the extent of the tails of the distribution, although the exact
percentile it corresponds to depends on the spatial resolution of the
simulation. Dilution due to expansion has been scaled out to highlight the
relative change of the distributions as evolution proceeds.

The evolution of the distribution of the decay field values (second row of
Figure~\ref{fig:stats}) clearly shows the onset of instability a little after
$t=100/m$, rapid spreading of the distribution due to exponential
amplification of seed inhomogeneities, and self-limiting of the growth by
non-linear interactions when the scaled value of the decay field becomes of
order unity. As the decay field perturbation grows and becomes non-linear, it
is drawing the energy from the zero mode of the inflaton (top row of
Figure~\ref{fig:stats}), reducing the amplitude of its oscillations, and
eventually forcing the inflaton to become strongly inhomogeneous as well due
to non-linear backreaction. We should note here that although the amplitude of
the coherent inflaton oscillation decays, it does not go away altogether, and
the left-over homogeneous mode will eventually come to dominate the universe
expansion \cite{Dufaux:2006ee}, as its equation of state is effectively that
of pressureless dust, and it dilutes slower than inhomogeneous components,
which have equation of state closer to relativistic one.

Of particular interest to us is the distribution of the total energy density,
shown in the third row of Figure~\ref{fig:stats}. It is clearly very
inhomogeneous, with peak densities easily exceeding ten times the average.
After a brief transient, it quickly settles to a nearly stationary
distribution, which appears to be highly non-Gaussian (and is in fact plotted
on a logarithmic scale). We will come back to this point after we inspect the
spatial picture of the energy density distribution inside the simulation box.

The bottom row of Figure~\ref{fig:stats} shows the evolution of the
gravitational potential. Despite total energy density being highly
inhomogeneous and having huge over-densities, the gravitational potential it
produces is rather small (at $10^{-3}$ level), and is further diluted away by
the expansion with near-relativistic equation of state. The maximal potential
well depth inside the simulation box is $2 \cdot 10^{-3}$, which is far too
small to form any primordial black holes. This result is in line with
observations of \cite{Suyama:2004mz,Suyama:2006sr}, although now we have a
more transparent diagnostic of black hole formation as we calculate the
gravitational potential directly.

The right panel of Figure~\ref{fig:stats} shows evolution of the field spectra
as the density plot in terms of time $t$ and wavenumber $k$, with dilution of
the fields due to expansion and overall power-law dependence on $k$ scaled
out. The spectra of total energy density $\rho_k$ and gravitational potential
$\Psi_k$ show a very clear simultaneous peak soon after instability develops,
which is sharply localized in both time and scale. Subsequent evolution of the
two differs, however. The peak power in energy density is evolving toward
higher $k$ and smaller scales, while the peak power in potential is evolving
to lower $k$, so the structure of potential wells is growing in spatial
extent. The understanding of what's going on will become more clear when we
look at evolution in real space, which is what we are going to discuss next.

\begin{figure*}
  \begin{center}
  \begin{tabular}{cc}
    $t=124/m$ & $t=256/m$ \smallskip\\
    \epsfig{file=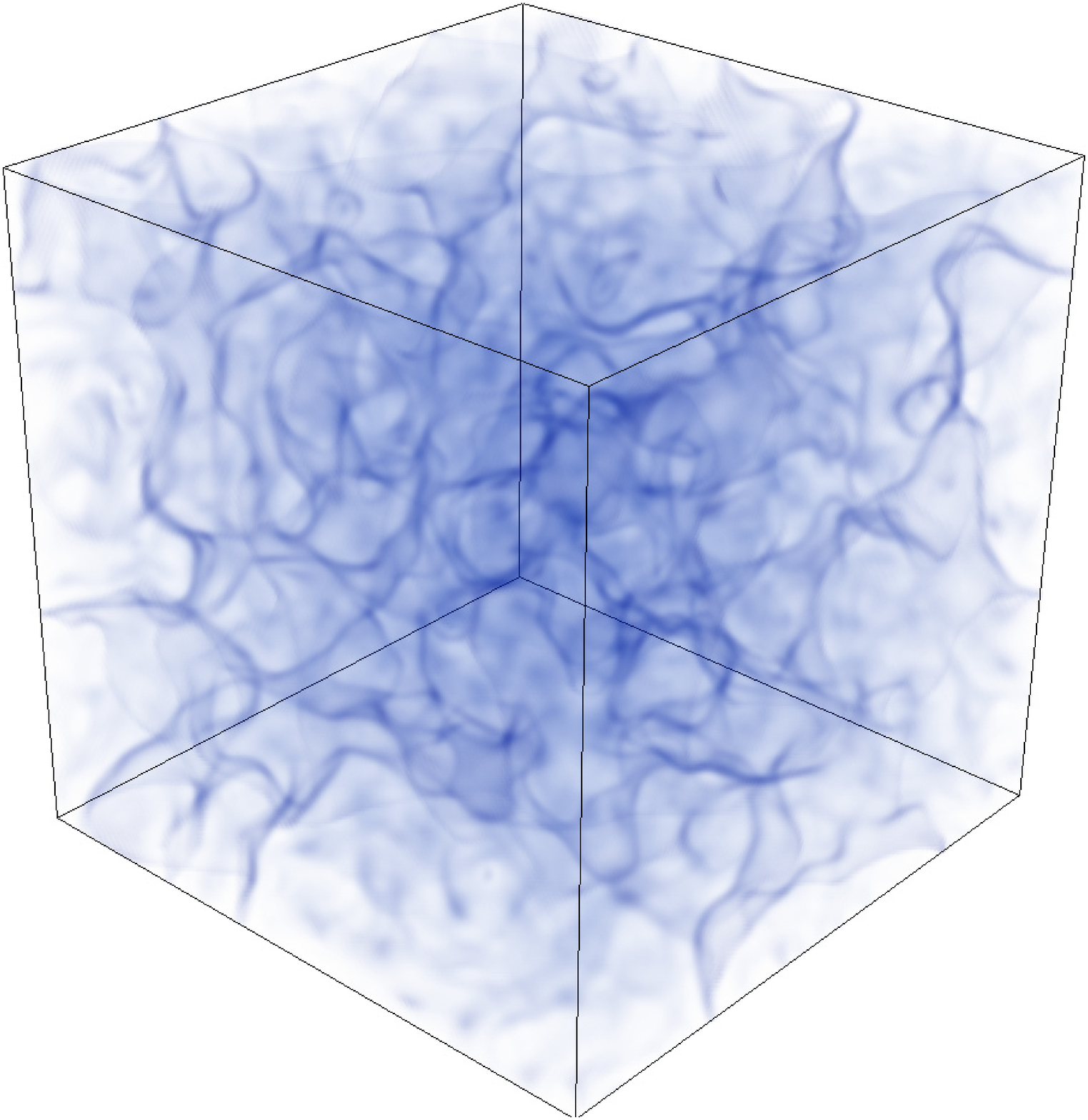, width=220pt} &
    \epsfig{file=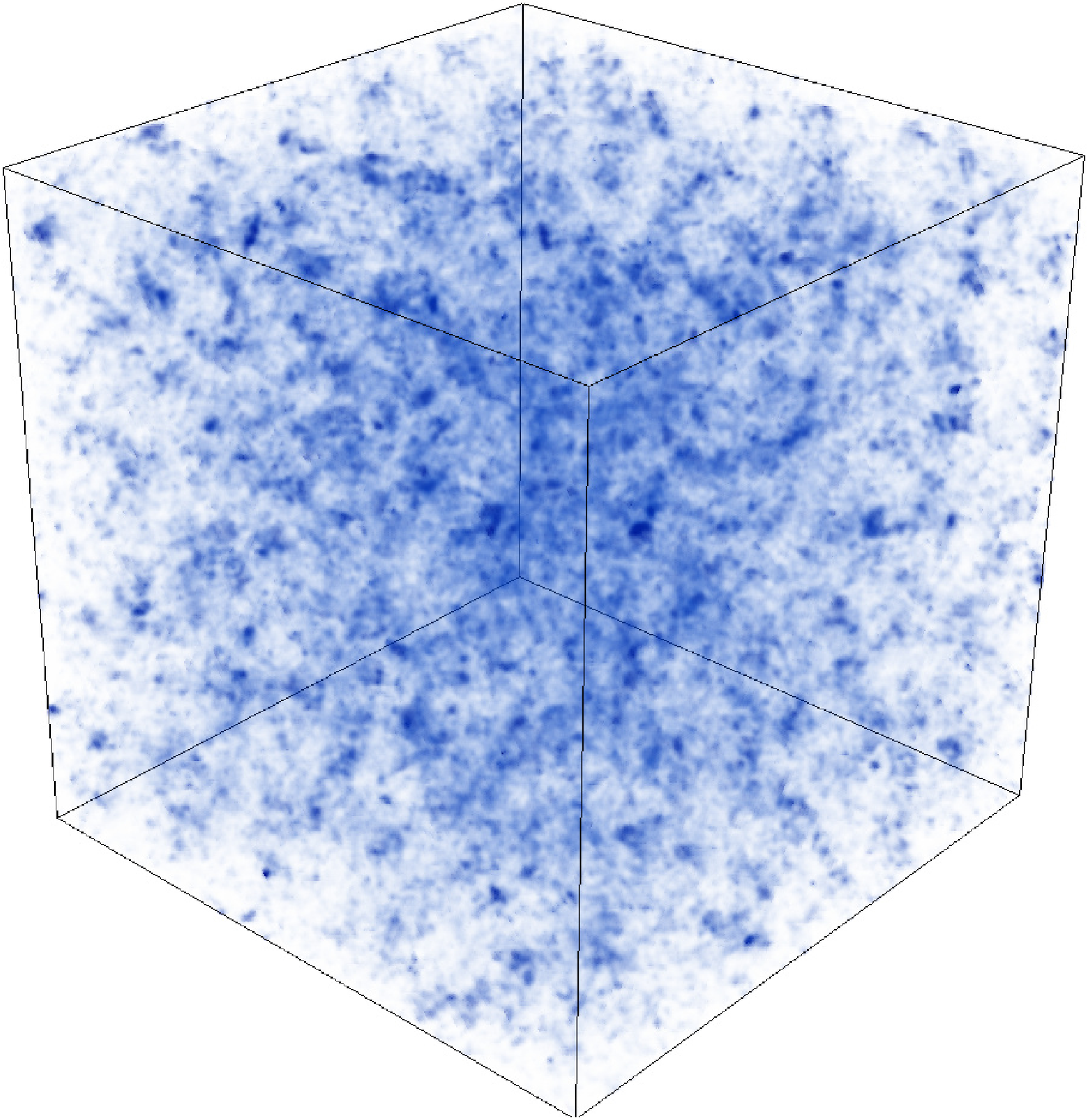, width=220pt} \\
    \epsfig{file=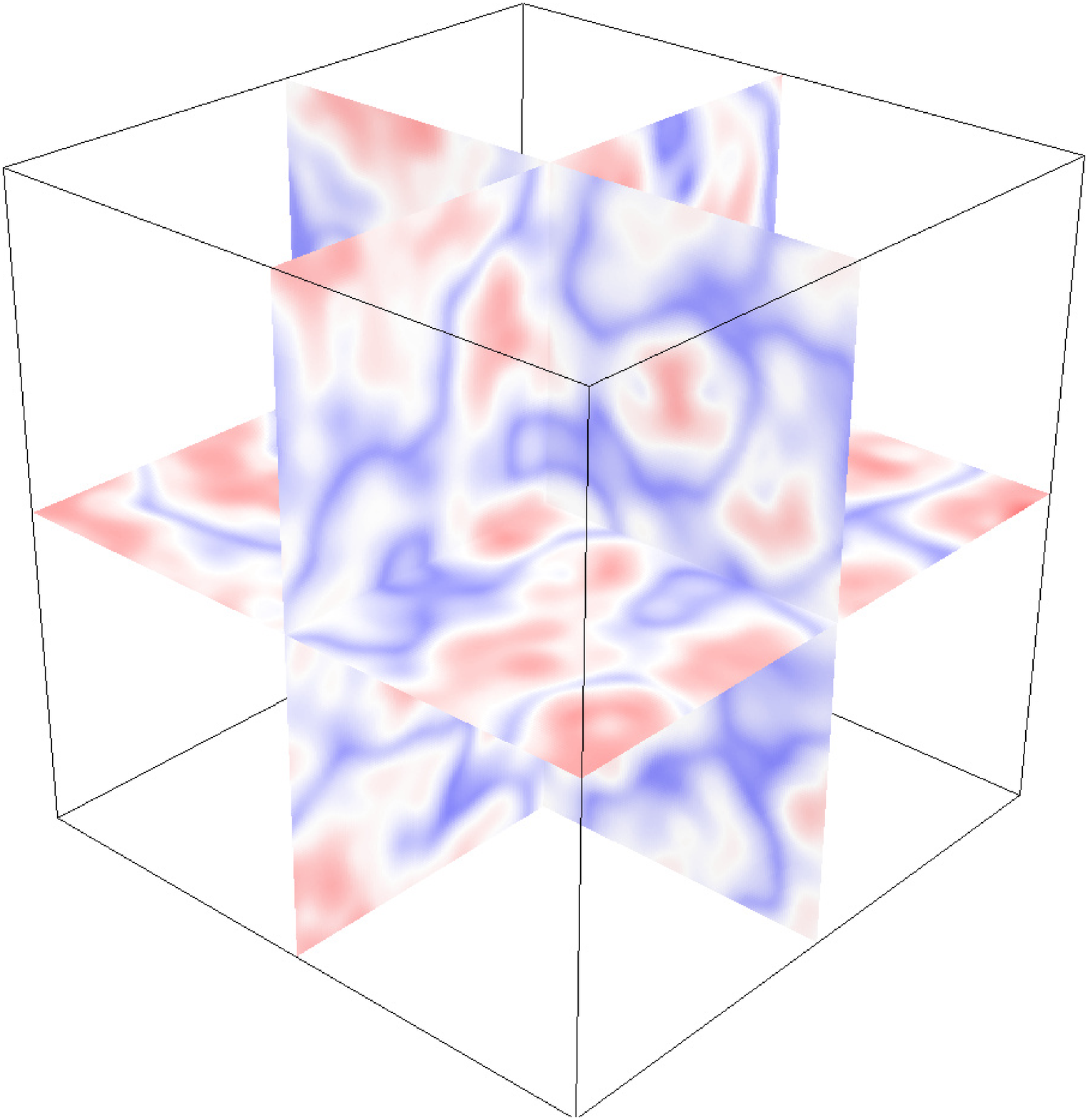, width=220pt} &
    \epsfig{file=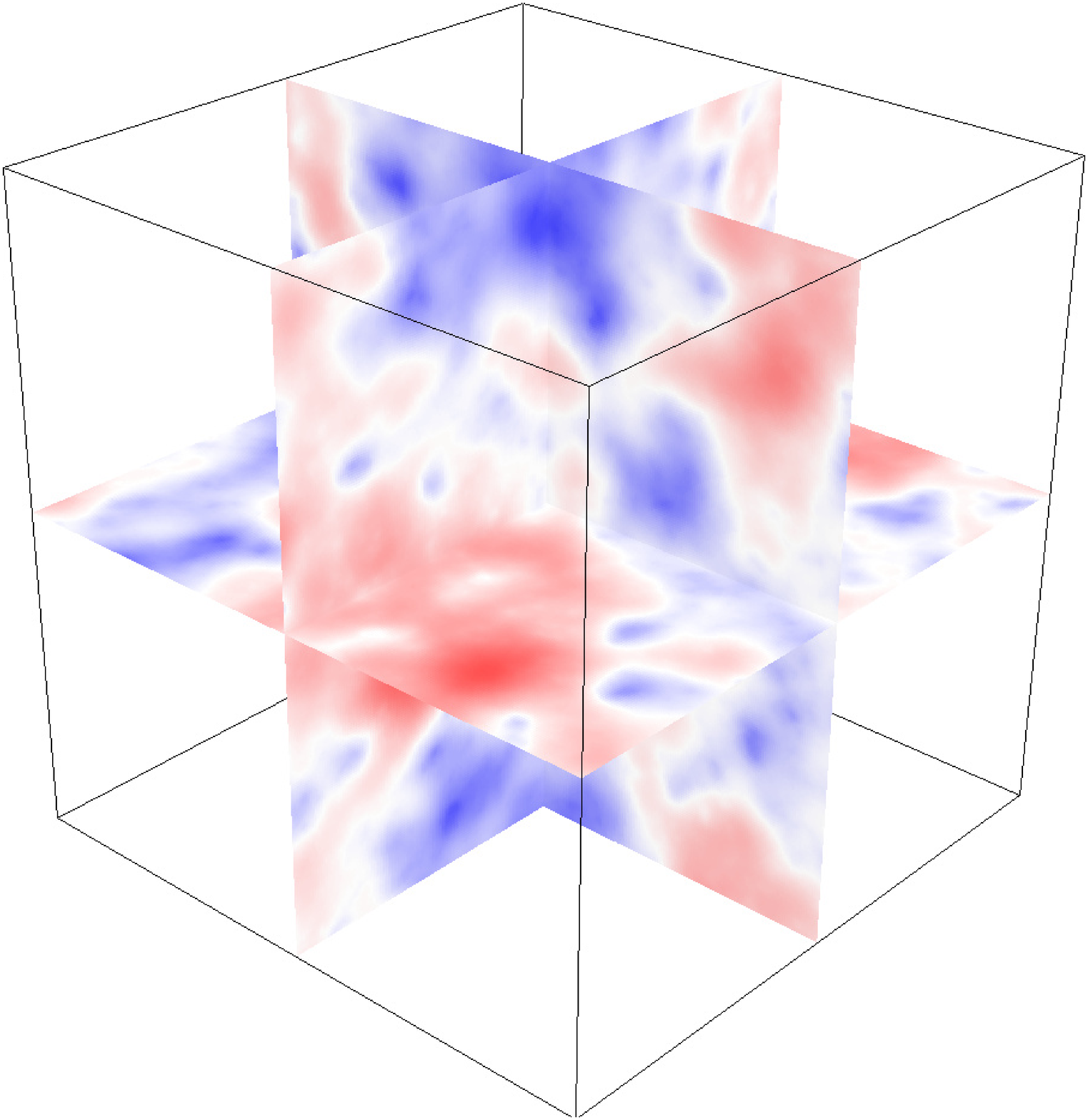, width=220pt} \\
    \epsfig{file=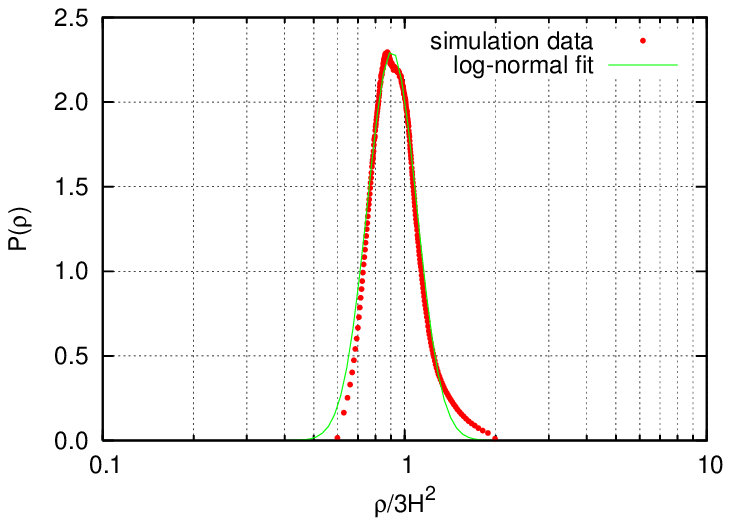, width=246pt} &
    \epsfig{file=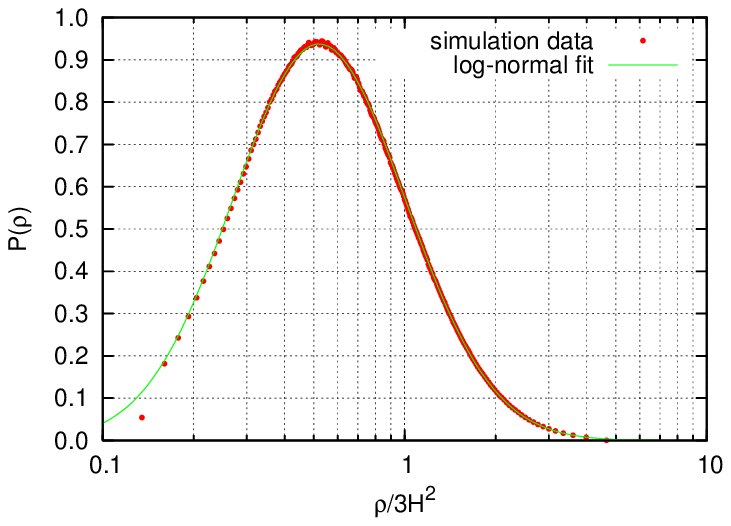, width=246pt} \vspace{-12pt}\\
  \end{tabular}
  \end{center}
  \caption{
    Volume rendering of total density $\rho$ (top row), gravitational
    potential $\Psi$ (second raw), and PDF of density values $\rho$ inside the
    simulation box soon after onset of instability (at $t=124/m$, left) and
    during subsequent evolution (at $t=256/m$, right). Animations for
    \href{http://www.sfu.ca/physics/cosmology/defrost/M2G22/volume-rho.avi}{density} and 
    \href{http://www.sfu.ca/physics/cosmology/defrost/M2G22/volume-PSI.avi}{potential}
    evolution are available at \url{http://www.sfu.ca/physics/cosmology/defrost/}.
  }
  \label{fig:rho}
\end{figure*}

Figure~\ref{fig:rho} shows three-dimensional volume renders of the contents of
the simulation box. As the fields $\phi$ and $\psi$ oscillate rapidly in a
standing wave pattern, quickly losing coherent phasing, the view of their
values is messy and not very enlightening, and we will omit it here. Much
more interesting is the picture of what is happening to the total energy
density, which is an adiabatic invariant for the oscillating fields. The top
row of Figure~\ref{fig:rho} shows density distribution inside the simulation
box soon after onset of instability (at $t=124/m$, left) and at the end of the
simulation (at $t=256/m$, right). Density field is shaded using logarithmic
color map with linear transparency ramp applied, so that only the peaks of the
density distribution are visible.

Immediately after the onset of instability, the density distribution in
Figure~\ref{fig:rho} (top left) looks like smoke is filling the box. What you
are seeing is actually the over-dense bubble walls forming a three-dimensional
foam-like structure that fills the box. Its origin is easy to understand if
one thinks about how seed inhomogeneities are amplified by instability. Broad
parametric resonance amplifies wavemodes in a certain band, effectively
serving as a low-pass filter (with a kernel that can be approximated
analytically) and sets the characteristic size of the structure which grows
out of the seed inhomogeneities. Original fluctuations are a Gaussian random
field, which already has the structure of peaks, ridges, and valleys imprinted
into it. The skeleton of this structure is essentially preserved unchanged as
the growth of inhomogeneities due to instability increases density contrast.
Once the density contrast becomes of order unity, non-linear evolution takes
over. This will happen to under-dense regions first, with repulsive
interaction term $g^2 \phi^2 \psi^2$ helping to evacuate the bubble interiors,
and pushing the matter density into the bubble walls, thus forming the
structure you see in Figure~\ref{fig:rho} (top left).

\begin{figure*}
  \begin{center}
  \begin{tabular}{cc}
    \epsfig{file=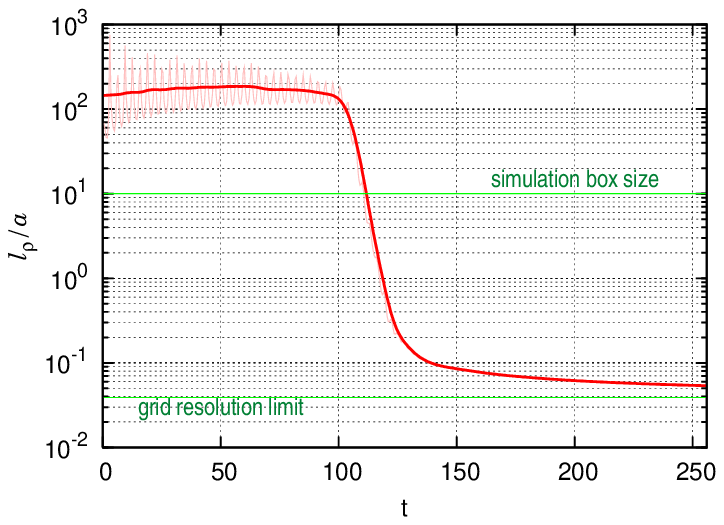, width=246pt} &
    \epsfig{file=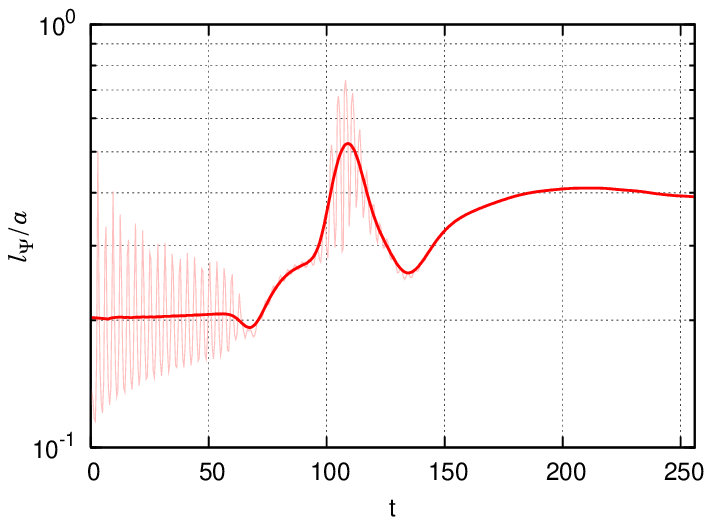, width=246pt}
  \end{tabular}
  \end{center}
  \vspace{-16pt}
  \caption{
    Evolution of characteristic structure size in total energy density (left) and gravitational potential (right).
  }
  \label{fig:scale}
\end{figure*}

The evolution does not stop at forming bubbles, however. The repulsive
interaction soon breaks the extended bubble walls into smaller localized
blobs, moving more or less freely inside the simulation box (the animation of
this process is available
\href{http://www.sfu.ca/physics/cosmology/defrost/M2G22/volume-rho.avi}{online}).
The final state is depicted in Figure~\ref{fig:rho} (top right), and persists
with little change for a long, long time. This state seems quite distinct from
thermal equilibrium, yet it is still long-lived and statistically simple in a
certain way, which is quite surprising. Even more surprisingly, the
distribution of values of total energy density $\rho$ in units of $H^2$
quickly becomes statistically stationary and after a brief transient tends to
a distribution with a probability density function shown in the lower right of
Figure~\ref{fig:rho}. It can be fitted with exceedingly high accuracy by a
lognormal distribution
\begin{equation}
  P(\rho)\, d\rho = \frac{1}{\sqrt{2\pi}\,\sigma}\, \exp\left[ - \frac{(\ln\rho - \mu)^2}{2 \sigma^2}\right] \frac{d\rho}{\rho}
\end{equation}
with one free parameter ($\sigma=0.6584$ or $\mu=-0.2197$), as the mean
$\bar\rho = \exp(\mu+\sigma^2/2)$ is unit-normalized by virtue of us scaling out
the expansion. The corresponding median is $e^\mu = 0.8028$. With statistical
errors of PDF estimator being what they are, the apparent lognormality of a
density distribution is undoubtedly not a mere coincidence, but must have a
explanation rooted in scalar field dynamics. Moreover, further simulations of
preheating models with different couplings and inflaton potentials seem to
suggest that lognormal distribution of density is a {\em universal feature} of
two-field preheating models. This observation presents a very interesting
theoretical puzzle, which will be explored in detail elsewhere
\cite{Frolov:preview}.

Although one-point distribution of total energy density $\rho$ quickly becomes
stationary as noted above, other quantities continue evolving on much longer
time scales. The blobs continue to fragment, and their characteristic size
slowly decreases with time. While this is obvious from visualizations, it can
be further quantified by introducing the (physical) correlation length of the
total energy density configuration
\begin{equation}
  \ell_\rho^2 \equiv \frac{\langle (a\rho)^2\rangle}{\langle(\nabla\!\rho)^2\rangle}.
\end{equation}
The evolution of the comoving correlation length $\ell_\rho/a$ is plotted in
the left panel of Figure~\ref{fig:scale}. Initially it is very large ($\gtrsim
100/m$), as the density field is nearly homogeneous. As instability develops
and structure forms, it abruptly drops to about $10^{-1}/m$, and then
continues to decrease, but much more slowly. Although the graph clearly shows
evolutionary trend in density correlation length $\ell_\rho/a$, actual numbers
should be taken with a grain of salt, as the density field does eventually
become fragmented on a scale close to spatial grid resolution (which for
$256^3$ grid we use would be reached around $t\sim 10^3/m$).

While it is known that thermalization after preheating might take a long time
\cite{Kofman:1995fi}, and there might be an intermediate scaling regime in the
evolution of the fields \cite{Micha:2002ey, Micha:2004bv}, the view of the
process from the real space presented here is strikingly simple. One would
think the field distributions will thermalize eventually, presumably forming a
homogeneous fluid-like state with Maxwellian distribution of particle
velocities. Thermalization {\em does not happen} in our simulations, which
lack necessary quantum effects. Exactly \textit{how} and \textit{when} it does
happen is an extremely interesting question, and the one which requires
further study.

Finally, let us discuss gravitational potential $\Psi$, which is sourced by
the evolving energy density distribution $\rho$, and is shown in the middle
row of Figure~\ref{fig:rho}. To make the structure more visible, we have opted
for a density plot on a three-slice through the simulation box rather than a
volume rendering. The color map shows positive potential values (corresponding
to under-dense regions) as shades of red, and negative potential values
(corresponding to over-dense regions) as shades of blue, blending into white
for zero potential value.

Immediately after the onset of instability, the gravitational potential in
Figure~\ref{fig:rho} (middle left) clearly traces the foam-like structure of
matter distribution. The isolated potential peaks (red) in the interior of the
bubbles are separated by extended potential valleys (blue) created by
over-dense bubble walls. The gravitational potential configuration is
asymmetric between positive and negative values, and is clearly non-Gaussian.
Subsequent evolution of the gravitational potential is rather interesting. As
bubble walls break into smaller and smaller blobs, the structure of the
gravitational potential does not follow suit. Instead, it begins to grow in
spatial extent (the animation is available
\href{http://www.sfu.ca/physics/cosmology/defrost/M2G22/volume-PSI.avi}{online}).
By the end of the simulation in Figure~\ref{fig:rho} (middle right), the size
of the structure in the gravitational potential spans almost the whole box.
The growth of structure can be quantified by introducing correlation length
$\ell_\Psi$ for gravitational potential the way we did for energy density
\begin{equation}
  \ell_\Psi^2 \equiv \frac{\langle (a\Psi)^2\rangle}{\langle(\nabla\!\Psi)^2\rangle}
    = -\frac{\langle 2\,\Psi^2\rangle}{\langle\rho\!\cdot\!\Psi\rangle}.
\end{equation}
Evolution of the comoving correlation length $\ell_\Psi/a$ is plotted in the
right panel of Figure~\ref{fig:scale}, with pale line tracing its
instantaneous value, and thick red line giving a running average (with Kaiser
window) over few oscillations. While comoving correlation length grows
overall, evolution shown in Figure~\ref{fig:scale} (right) still represents an
initial transient, and the long-time asymptotic behaviour is not reached in
the simulation reported here. Further investigation shows that correlation
length $\ell_\Psi$ continues to grow faster than comoving size, but not quite
as fast as the horizon size. Eventually one might even have to worry about it
outgrowing the finite simulation box size, but that would take a very long
time, and is not reached by our simulations.

\section{Conclusions}\label{sec:concl}

This paper presents a new numerical code I developed for simulating preheating
of the Universe after the end of the inflation, which I call DEFROST. It is
small (about 600 lines of Fortran code), fast, easy to modify, and is fully
instrumented for 3D visualizations (using
\href{http://wci.llnl.gov/codes/visit/}{LLNL's VisIt}, for example).
The source code is available for download online at
\url{http://www.sfu.ca/physics/cosmology/defrost/}
and is distributed under the terms of GNU Public License. While the main
design goal of DEFROST has been the accuracy of the simulations, performance
of the solver has also been significantly improved compared to
LATTICEEASY~\cite{Felder:2000hq}, which is the most mature and widely used
reheating code publicly available today.

As a result of all the optimizations (and a bit of black magic), DEFROST
outperforms LATTICEEASY by about a factor of four in raw PDE solver speed (for
2 fields on a $256^3$ grid in double precision on a dual Xeon 5160 machine)
while using more accurate (and more expensive) discretization. If one takes
into account the time spent on analysis of the results, the difference is even
larger, as FFTW libraries used by DEFROST are vastly faster than FFT routines
shipped with LATTICEEASY (especially on multi-processor machines). The
speed-up offered by DEFROST is so significant that the studies done a few
years ago on a big parallel cluster \cite{Podolsky:2005bw, Dufaux:2006ee}
using MPI version of LATTICEEASY \cite{Felder:2007nz} can now be carried out
on a single fast workstation. The planned MPI version of DEFROST should be
able to push the accessible simulation size over the $1024^3$ barrier,
provided the code scales well.

The code was tested on a number of chaotic inflation models which end via
parametric resonance. In this paper, we report the simulations of the simplest
two-field preheating model with massive inflaton and quartic coupling to decay
field (\ref{eq:model}). We reproduce the previously published numerical
results for this model \cite{Podolsky:2005bw, Felder:2006cc} (and the ones for
trilinear coupling \cite{Dufaux:2006ee}, which we will not discuss here,
although our simulation data and results for that model are available
\href{http://www.sfu.ca/physics/cosmology/defrost/M2S12L4}{online} as well).
We further investigate the dynamics of the scalar field evolution in these
preheating models, taking advantage of advanced visualization and analysis
capabilities DEFROST offers. In particular, we study the behaviour of energy
density distribution and scalar gravitational potential during preheating,
something which has not been looked at closely before. Our main science
results are summarized by two observations, both novel and quite surprising.

First, the evolving scalar fields quickly end up in a simple state, which,
although highly inhomogeneous, appears to have a certain universality to it.
In this state, the one-point distribution function of total energy density is
nearly stationary (apart for the overall dilution due to expansion), and is
described by a lognormal distribution for {\em all two-field parametric
resonance preheating models we tried so far}, namely the ones described by
interaction potentials
\begin{itemize}
  \item $V(\phi,\psi) = \frac{1}{2}\, m^2 \phi^2 + \frac{1}{2}\, g^2 \phi^2 \psi^2$,
  \item $V(\phi,\psi) = \frac{1}{2}\, m^2 \phi^2 + \frac{1}{2}\, \sigma \phi \psi^2 + \frac{1}{4}\, \lambda \psi^4$,
  \item $V(\phi,\psi) = \frac{1}{4}\, \lambda \phi^4 + \frac{1}{2}\, g^2 \phi^2 \psi^2$,
  \item $V(\phi,\psi) = \frac{1}{4}\, \lambda (\phi^2 + \psi^2)^2$.
\end{itemize}
This is true even if distributions of field values or other correlators might
still be evolving, and appears to be a very general statement about random
scalar fields one encounters in preheating. It is tempting to attribute this
state to scalar field turbulence \cite{Micha:2002ey, Micha:2004bv}, especially
since lognormal density distributions are known to occur in supersonic
isothermal turbulence in hydrodynamics \cite{Nordlund:1998wj}. We do not see
obvious signs of thermalization, even if the simulations are run for a time
much longer than the dynamical timescale of the problem (the longest done so
far for massive inflaton is $2^{12}/m$ corresponding to five $e$-folds since
the end of inflation; this is limited mainly by my patience rather than the
code stability).

Second, less general but still amusing, is the observation that the
small-scale structure in the gravitational potential can grow faster than
comoving box expands. It is not quite clear whether the reason for it
happening is kinematic or dynamical in nature. As we neglected gravitational
interactions in our simulations, the only thing that can cause the structure
to grow is the interaction between scalar fields themselves. In our preheating
model it is repulsive; yet the structure still grows! Although one might
suspect that any inhomogeneity in gravitational potential on sub-horizon
scales would probably get washed away by subsequent evolution (and is too
small to form primordial black holes), this effect still might have some
interesting cosmological consequences.

All in all, we find that the picture of preheating dynamics is simpler in real
space than what it looks like in particle representation. The final stage of
preheating, with growing structure and lognormal density distribution, eerily
reminds one of large scale structure formation in later cosmology (although of
course it occurs on vastly smaller scales and is driven by completely
different physics). Perhaps the analytical methods developed for the latter
\cite{Bardeen:1985tr, Bernardeau:1994aq, Bond:1995yt} could be fruitfully
applied to preheating as well. This is what we intend to explore next.

\section*{Acknowledgments}

This work was supported by the Natural Sciences and Engineering Research
Council of Canada under Discovery Grants program. Numerical computations were
done in part on Sunnyvale cluster at Canadian Institute for Theoretical
Astrophysics. The author is grateful for hospitality during his stays at CITA,
where some portion of this work has been carried out, and would like to thank
Lev Kofman, Andrei Linde, Gary Felder, and Mustafa Amin for helpful
discussions of preheating, and Chris Matzner for bringing the subject of
isothermal supersonic turbulence to my attention.


\newpage

\end{document}